\documentclass{aa}   
\usepackage{graphicx}
\usepackage{txfonts}
\usepackage{xcolor}
\begin{document} 
%-----------------------------------------------------------------------

\title{The VMC survey -- XXXVIII.}
\subtitle{Proper motion of the Magellanic Bridge\thanks{Based on observations made with VISTA at the La Silla Paranal Observatory under programme ID 179.B-2003.}}

\author{Thomas~Schmidt\inst{1},
        Maria-Rosa~L.~Cioni\inst{1},
        Florian~Niederhofer\inst{1},
        Kenji~Bekki\inst{2},
        Cameron~P.~M.~Bell\inst{1},
        Richard~de~Grijs\inst{3,4,5},
        Jonathan~Diaz\inst{2},
        Dalal~El~Youssoufi\inst{1}, 
        Jim~Emerson\inst{7},
        Martin~A.~T.~Groenewegen\inst{7},
        Valentin~D.~Ivanov\inst{8},
        Gal~Matijevic\inst{1},
        Joana~M.~Oliveira\inst{9},
        Monika~G.~Petr-Gotzens\inst{8,10},
        Anna~B.~A.~Queiroz\inst{1},
        Vincenzo~Ripepi\inst{11},
        Jacco~Th.~van~Loon\inst{9}
        }

\institute{Leibniz-Institut für Astrophysik Potsdam (AIP), An der Sternwarte 16, D-14482 Potsdam, Germany\\
           \email{tschmidt@aip.de}
           \and
           ICRAR, M468, The University of Western Australia, 35 Stirling Hwy, Crawley Western Australia 6009, Australia
           \and
           Department of Physics and Astronomy, Macquarie University, Balaclava Road, Sydney NSW 2109, Australia
           \and
           Centre for Astronomy, Astrophysics and Astrophotonics, Macquarie University, Balaclava Road, Sydney NSW 2109, Australia
           \and
           International Space Science Institute--Beijing, 1 Nanertiao, Zhongguancun, Hai Dian District, Beijing 100190, China
           \and
           Astronomy Unit, School of Physics and Astronomy, Queen Mary University of London, Mile End Road, London E1 4NS, UK
           \and
           Koninklijke Sterrenwacht van Belgi\"e, Ringlaan 3, B–1180 Brussels, Belgium
           \and
           European Southern Observatory, Karl-Schwarzschild-Str. 2, D-85748 Garching bei München, Germany
           \and
           Lennard-Jones Laboratories, School of Chemical and Physical Sciences, Keele University, ST5 5BG, UK
           \and
           Universitäts-Sternwarte, Ludwig-Maximilians-Universität München, Scheinerstr 1, D-81679 München, Germany
           \and
           INAF-Osservatorio Astronomico di Capodimonte, via Moiariello 16, I-80131 Naples, Italy
           \\
           %    \email{XXX@XX.COM}
           %    \thanks{The university of heaven temporarily does not
           %            accept e-mails}
            }

\date{submitted Jan 10, 2020; accepted Jun 02, 2020}
\titlerunning{Proper motion maps of the Magellanic Bridge}
\authorrunning{Schmidt et al.}
%-----------------------------------------------------------------------

% \abstract{}{}{}{}{} 
% 5 {} token are mandatory
 
  \abstract
   {The Magellanic Clouds are a nearby pair of interacting dwarf galaxies and satellites of the Milky Way. Studying their kinematic properties is essential to understanding their origin and dynamical evolution. They have prominent tidal features and the kinematics of these features can give hints about the formation of tidal dwarfs, galaxy merging and the stripping of gas. In addition they are an example of dwarf galaxies that are in the process of merging with a massive galaxy.}
   {The goal of this study is to investigate the kinematics of the Magellanic Bridge, a tidal feature connecting the Magellanic Clouds, using stellar proper motions to understand their most recent interaction.}
   {We calculated proper motions based on multi-epoch $K_{s}$-band aperture photometry, which were obtained with the Visible and Infrared Survey Telescope for Astronomy (VISTA), spanning a time of 1-3 yr, and we compared them with $Gaia$~Data~Release~2~(DR2) proper motions. We tested two methods for removing Milky Way foreground stars using $Gaia$~DR2 parallaxes in combination with VISTA photometry or using distances based on Bayesian inference.}
   {We obtained proper motions for a total of 576,411 unique sources over an area of $23$~deg$^{2}$ covering the Magellanic Bridge including mainly Milky Way foreground stars, background galaxies, and a small population of possible Magellanic Bridge stars ($<$15,000), which mostly consist of giant stars with $11.0<K_{s}<19.5$~mag. The first proper motion measurement of the Magellanic Bridge centre is $1.80\pm0.25$~mas~yr$^{-1}$ in right ascension and $-0.72\pm0.13$~mas~yr$^{-1}$ in declination. The proper motion measurements of stars along the Magellanic Bridge from the VISTA survey of the Magellanic Cloud system (VMC) and $Gaia$~DR2 data confirm a flow motion from the Small to the Large Magellanic Cloud. This flow can now be measured all across the entire length of the Magellanic Bridge.}
   {Our measurements indicate that the Magellanic Bridge is stretching. By converting the proper motions to tangential velocities, we obtain $\sim$110~km~s$^{-1}$ in the plane of the sky. Therefore it would take a star roughly 177~Myr to cross the Magellanic Bridge.}


   \maketitle
%________________________________________________________________
\section{Introduction} \label{Introduction}

\begin{figure*}
   \centering
   \includegraphics[width=\textwidth]{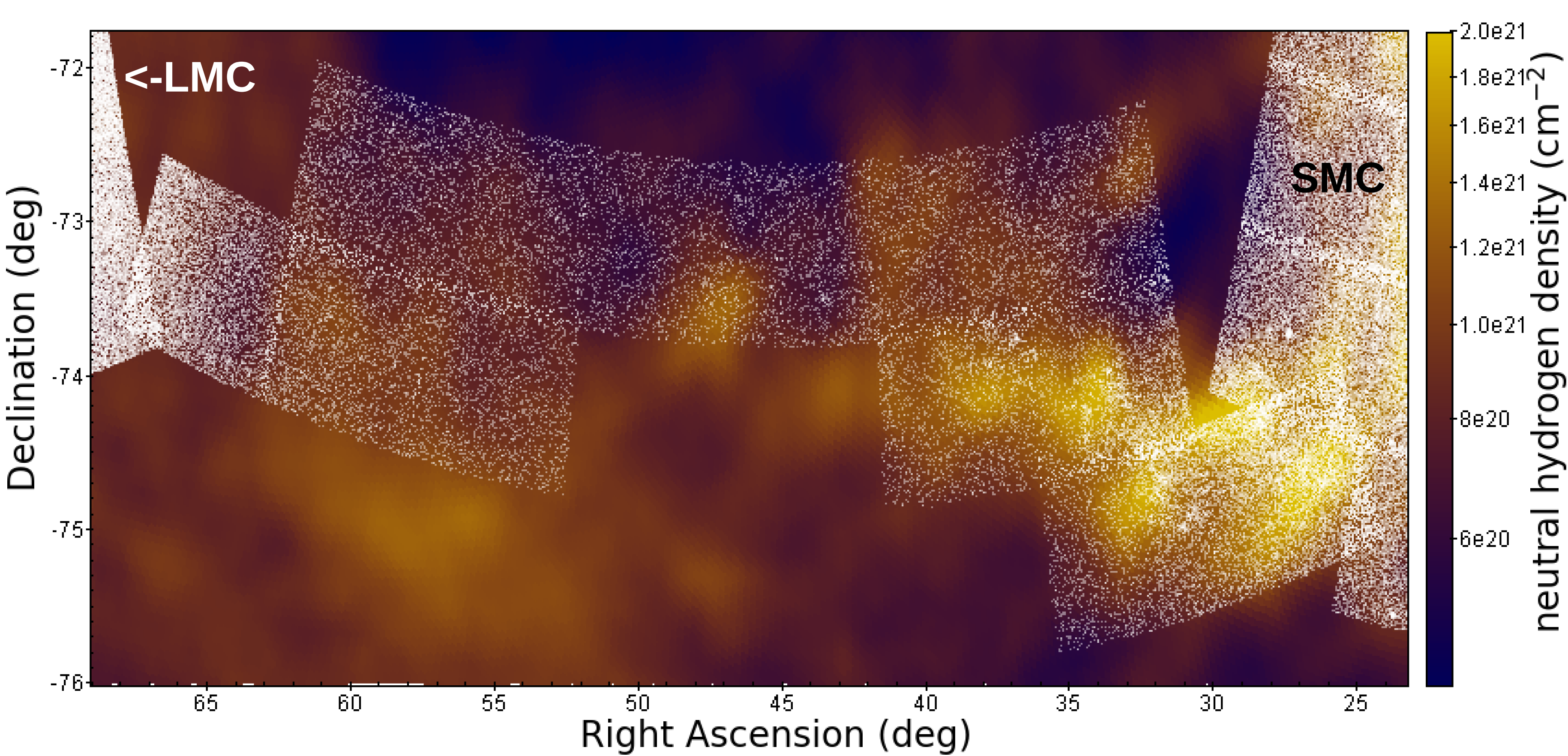}
   \caption{Distribution of Magellanic Bridge stars from the VMC-$Gaia$~DR2 sample (white) superimposed on the distribution of \ion{H}{I} gas \citep{HI4PI2016}. The rectangular structures represent the shape of the VMC tiles.}
              \label{Figure1}%
    \end{figure*}

The Magellanic Clouds (MCs) are two satellite dwarf galaxies ($10^{9}-10^{11}$M$_{\odot}$) of the Milky Way (MW) and an example of an early stage of minor mergers. A minor merger is a process of merging a significantly smaller galaxy with a more massive galaxy (mass ratio $\sim$10:1). It has been suggested that the MCs are on their first infall into the MW \cite[e.g.][]{Besla2007} and that in the future there will be two possible merging processes: the merging of the Small Magellanic Cloud (SMC) into the Large Magellanic Cloud (LMC) \citep[in $\sim$2~Gyr,][]{Besla2016} and the merging of the MCs into the MW \cite[in $\sim$3~Gyr,][]{Cautun2019}.
Both the LMC and SMC are classified as dwarf irregular galaxies. The LMC resembles a nearly face-on spiral and is around ten times more massive than the SMC \citep[e.g.][]{Penarrubia2016, Bekki2009}, also the SMC is significantly elongated along the line-of-sight \citep[25-30 kpc; e.g.][]{Ripepi2017, Subramanian2009}. 
It is unclear how many interactions the MCs have had in their past. Dynamical simulations suggest a minimum of two interactions \citep[e.g.][]{Besla2016, Pearson2018}. It is also unclear whether the LMC is bound to the MW and if the SMC is bound to the LMC (e.g. Gonzalez \& Padilla 2016). \citet{Besla2016} suggest that the LMC and the SMC have been bound for $\sim$6.3~Gyr. A major factor in these uncertainties is the total mass of the MCs (e.g. $m_{\text{LMC}}=(1.7 \pm 0.7)\times 10^{10} \textit{M}_{\odot}$ within $8.7$ kpc, \citealt{vanderMare2014}; $m_{\text{SMC}}=2.4\times 10^{9}$M$_{\odot}$, \citealt{Stanimirovic2004}). Other studies suggest much larger masses for the LMC \citep[e.g. $m_{\text{LMC}}$=$1.38$$\times 10^{11} \textit{M}_{\odot}$,][]{Erkal2019}. Such a difference in mass has a significant impact on our understanding of past interactions. Dynamical simulations with lower LMC masses \citep[e.g.][]{Bekki2007, Besla2013} suggest that the last direct interaction between both dwarf galaxies occurred about 200 Myr ago. More recent results from \citet{Zivick2019} claim this interaction to be more recent (147$\pm$33 Myr ago). Additional constraints can be obtained by studying substructures, which are directly associated with the last interaction. The substructures in the Magellanic system that are probably associated with this interaction are as follows: a bar offset from the centre of the LMC disc, the 30 Doradus starburst region, which was created by a massive inflow of gas \citep[e.g.][]{Bekki2013}, an extended wing of the SMC towards the LMC, and a bridge of neutral hydrogen gas (\ion{H}{I}) and stars connecting the LMC with the SMC.

The Magellanic Bridge was first discovered by \cite{Hindman1963} from an over-density of \ion{H}{I} connecting the MCs. Subsequent studies have shown that the Magellanic Bridge was probably formed by tidal forces stripping gas mostly from the SMC \citep{Murai1980,Gardiner1996}. Both studies, based on numerical simulations, implied a recent burst of star formation including the presence of early-type stars. This suggestion is supported by observations \citep[e.g.][]{Irwin1985, Dufton2008, Carrera2017} indicating that the Magellanic Bridge was formed by the last interaction between the LMC and SMC. These studies searched for specific tracers such as a young stellar population ($<$~$300$~Myr old) which is believed to have been formed in situ \citep{Irwin1985}. A recent study of the 3D kinematics of gas in the SMC \citep{Murray2019} finds that stripped SMC stars show a radial velocity gradient in agreement with the \ion{H}{I} radial velocity field. An older population of stars was expected to be present as well since tidal forces have similar effects on stars and gas. Later observational studies \citep[e.g.][]{Bagheri2013,Noel2013} presented evidence of this older population in the Magellanic Bridge. \citet[][]{Bagheri2013} found that the ages of red giant branch (RGB) and asymptotic giant branch stars in the central Bridge region are likely to range from 400~Myr to 5~Gyr. This age range implies that these stars did not form in situ and were stripped into the Magellanic Bridge by tidal forces during the last interaction between the LMC and SMC. This implication is supported by dynamical simulations \citep[e.g.][]{Guglielmo2014}. However, the first spectroscopic evidence of a stellar population older than 1 Gyr between the MCs was presented by \citet{Carrera2017}. The metallicity of this population suggests its origin to be more likely in the outer regions of the SMC. Stars that formed from the stripped gas conversely were shown to have metal abundances more consistent with having been formed in situ \citep{Dufton2008}.

The Magellanic Bridge covers a large area of the sky between the MCs which is at least twice the size of the 23~deg$^{2}$ area covered by the VISTA survey of the Magellanic Clouds system \citep[VMC;][]{Cioni2011}.
The overlap between the VMC tiles and the \ion{H}{I}~gas is shown in Fig.~\ref{Figure1}. Also shown are some of the neighbouring VMC tiles of the LMC (left) and SMC (right). The VMC-Bridge tiles follow the gas-rich Wing of the SMC, cover the densest gas region in the centre and connect to the LMC.
Distances based on classical Cepheids indicate that the Magellanic Bridge extends $\sim$20~kpc in the plane of the sky from the northeast (LMC side) to the south-west (SMC side) and $\sim$10~kpc along the line of sight \citep[e.g.][]{JacyszynDobrzeniecka2016}. Many studies focused on the western side of the Magellanic Bridge, which is east of the SMC Wing, where the stellar density is high compared to the central Bridge regions. A second bridge was later claimed by \citet{Belokurov2017} using $Gaia$~Data~Release 1 (DR1) data. This candidate bridge is composed of RR Lyrae stars and is connecting the LMC and SMC in an arc southwards of the \ion{H}{I} bridge. It was named the 'old bridge', since RR Lyrae stars are old compared to the young main-sequence stars of the first bridge. The existence of this old bridge is less clear according to \citet{JacyszynDobrzeniecka2020} as it could be explained by the overlap of the LMC and SMC halos. Further investigations are needed and more detailed kinematics could resolve this issue.

Proper motion measurements are one step towards obtaining detailed kinematics of the MCs. They add two key components of the three-dimensional velocity necessary for a full understanding of their kinematics. Previous proper motion studies \citep[e.g.][]{Kallivayalil2006,Costa2009,Costa2011,vanderMarel2016} presented highly accurate proper motion measurements of the main bodies of the LMC and SMC and were focused on regions with relatively high stellar densities. Those studies support the idea that the MCs are moving together, while also showing that the centres of both galaxies are currently moving apart. In this study, we focus on the proper motion measurements between the MCs, where the Magellanic Bridge is located, and concentrate on the kinematics of this tidal structure. The first proper motion maps of the Magellanic Bridge indicating a motion of stars from the SMC towards the LMC were presented by \citet{Schmidt2018} followed by a kinematic analysis based on $Gaia$~DR2 and Hubble Space Telescope (HST) data by \citet{Zivick2019}. Both studies were limited by residual MW foreground stars in the centre of the Magellanic Bridge.

Herewith we present new proper motion measurements based on data from the VMC. We found that a significant MW foreground removal is needed when studying regions as sparsely populated as the central parts of the Magellanic Bridge. Applying the same selection criteria presented in previous studies of the Clouds using $Gaia$~DR2 data \citep[e.g.][]{Helmi2018, Zivick2019,Vasiliev2018} proved not to be applicable to our specific case since they either removed too few or too many sources. In this study we introduced two methods. The first method combines a colour-magnitude selection of stars based on VMC data with $Gaia$~DR2 parallaxes (see Sect.~\ref{method1}). The second method employs a selection based on distances computed from $Gaia$~DR2 stellar parameters. Our novel methods significantly increase the fraction of reliable sources by efficiently removing foreground stars without being too restrictive. These improvements enable us to study the kinematics of more sparsely populated regions such as the Magellanic Bridge centre or the outskirts of the LMC and significantly increase the spatial resolution of our proper motion maps. In upcoming works, we will explore the outer regions of the LMC covered by the VMC survey with these methods.

We organise the paper as follows. Section~\ref{Observations} describes the VMC observations used in this study. Section~\ref{Analysis} presents our data analysis and the methods used. It includes the VMC and $Gaia$~DR2 data selections, as well as the handling of the MW foreground stars and introduces our use of distances based on Bayesian inference. In Section~\ref{results} we compare the two methods of removing foreground MW stars. The comparison is followed by the resulting proper motion measurements and illustrated in proper motion maps obtained from both methods. In Section~\ref{Discussion} we discuss our results and we conclude the paper in Section~\ref{Conclusion}.

\section{Observations} \label{Observations}

    %_____________________________________________________________
    %--- VMC fields --- 
    \begin{table*}[]
    \caption{$K_{s}$ band Observations of the VMC Bridge tiles.}
    \label{Table1}
    \small
    \begin{tabular}{cccrccccc}
    \hline\hline
    Tile & Right ascension\tablefootmark{a} & Declination\tablefootmark{a} & Position angle\tablefootmark{b} & Epochs & Time baseline & FWHM\tablefootmark{c} & Airmass\tablefootmark{c} & Sensitivity \tablefootmark{c,d} \\
    & (h:m:s) & ($^\circ$:$^\prime$:$^{\prime\prime}$) & (deg) & & (day) & (arcsec) & &(mag)\\
    \hline
    BRI 1\_2 & 01:49:51.432 & $-$74:43:25.320 & $-$16.8805 & 10 & 1124 & 0.94$\pm$0.06 & 1.65$\pm$0.09 & 19.81 \\
    BRI 1\_3 & 02:11:34.464 & $-$75:05:00.960 & $-$11.6612 & 11 & 1084 & 1.00$\pm$0.09 & 1.61$\pm$0.03 & 19.85 \\
    BRI 2\_3 & 02:14:46.584 & $-$74:00:47.520 & $-$10.8627 & 12 & 1072 & 0.99$\pm$0.10 & 1.63$\pm$0.06 & 20.00 \\
    BRI 2\_4 & 02:35:28.440 & $-$74:13:18.840 & $-$5.8932 & 11 & 745 & 0.94$\pm$0.10 & 1.57$\pm$0.02 & 19.93 \\
    BRI 2\_7 & 03:39:50.712 & $-$74:04:51.240 & +9.5439 & 11 & 780 & 0.93$\pm$0.08 & 1.58$\pm$0.05 & 19.90 \\
    BRI 2\_8 & 04:00:21.072 & $-$73:46:37.560 & +14.4905 & 12 & 1056 & 0.90$\pm$0.09 & 1.57$\pm$0.04 & 19.96 \\
    BRI 2\_9 & 04:19:21.528 & $-$73:22:10.560 & +19.0897 & 10 & 786 & 0.92$\pm$0.07 & 1.58$\pm$0.06 & 19.83 \\
    BRI 3\_3 & 02:17:36.600 & $-$72:56:20.400 & $-$10.2104 & 11 & 1280 & 0.96$\pm$0.08 & 1.57$\pm$0.07 & 19.94 \\
    BRI 3\_4 & 02:37:26.016 & $-$73:08:16.080 & $-$5.4372 & 11 & 674 & 0.94$\pm$0.07 & 1.55$\pm$0.04 & 19.96 \\
    BRI 3\_5 & 02:57:33.288 & $-$73:12:52.200 & $-$0.5877 & 11 & 635 & 0.96$\pm$0.08 & 1.58$\pm$0.04 & 19.96 \\
    BRI 3\_6 & 03:17:45.000 & $-$73:10:02.640 & +4.2769 & 11 & 675 & 0.95$\pm$0.10 & 1.58$\pm$0.05 & 19.97 \\
    BRI 3\_7 & 03:37:39.240 & $-$72:59:54.600 & +9.0465 & 12 & 657 & 0.97$\pm$0.08 & 1.57$\pm$0.04 & 19.99 \\
    BRI 3\_8 & 03:57:04.968 & $-$72:42:31.680 & +13.7448 & 9 & 1088 & 0.98$\pm$0.12 & 1.57$\pm$0.06 & 19.91 \\
    \hline
    \end{tabular}\\
    \tablefoottext{a}{Coordinates of the VMC tile centres.}\\
    \tablefoottext{b}{Orientation of the VMC tiles, defined to increase from north to east.}\\
    \tablefoottext{c}{Average of all used epochs.}\\
    \tablefoottext{d}{For sources with photometric uncertainty <0.1 mag.}
    \end{table*}

Data presented in this study are taken from the VMC survey \citep[][]{Cioni2011}. The VMC survey started acquiring data in November 2009 and observations were completed in October 2018. The survey consists of multi-epoch near-infrared images in the $Y$, $J$, and $K_{s}$ filters of 110 overlapping tiles across the Magellanic system: 68 covering the LMC, 27 for the SMC, 13 for the Magellanic Bridge, and two for the Stream components. Each tile covers 1.77~deg$^2$ on the sky, consisting of 1.50~deg$^2$ with full and 0.27~deg$^2$ with half the total exposure time. In this study, we focus on the Magellanic Bridge tiles. The distribution of these tiles can be seen in Fig.~\ref{Figure1}, where they are superimposed on the distribution of \ion{H}{I} \citep{HI4PI2016}. 

The observations were obtained with the VISTA Camera (VIRCAM) on the Visible and Infrared Survey Telescope for Astronomy\footnote{http://www.vista.ac.uk} \citep[VISTA,][]{Sutherland2015} operated by the European Southern Observatory (ESO). VIRCAM is a near-infrared imaging camera composed of 16 VIRGO HgCdTe detectors. Each detector covers an area of 0.0372~deg$^2$ with an average spatial resolution of 0.339$^{\prime\prime}$~px$^{-1}$. The individual images from the 16 detectors form a VISTA pawprint that covers 0.6~deg$^2$ not including the gaps between the detectors. A mosaic of 6 pawprints was used to cover a continuous area filling the gaps between individual detectors. This arrangement forms a VMC tile. The individual detector integration time (DIT) for a $K_{s}$-band exposure was 5s. Taking 5 jitters and 15 repetitions into account this adds up to 750 seconds per tile and epoch. However, in a single pawprint each pixel is exposed on average for 375s per tile. Images were processed using the VISTA Data Flow System pipeline \citep[VDFS v1.3,][]{Emerson2006} at the Cambridge Astronomy Survey Unit\footnote{http://casu.ast.cam.ac.uk} (CASU) and stored in the VISTA Science Archive\footnote{http://horus.roe.ac.uk/vsa} \citep[VSA,][]{Cross2012}. There are at least 11 epochs at $K_{s}$ of this type (deep) and two epochs with half the exposure time (shallow). Exposure times in the $Y$ and $J$ bands as well as additional parameters of the survey are described in detail by \citet{Cioni2011}.
The catalogues provided by the VSA contain both aperture and point-spread-function (PSF) photometry. Their magnitudes have been calibrated as explained by \citet{Gonzalez2018} and result in an accuracy of better than 0.02~mag in $YJK_{s}$. This study uses the aperture photometry data because of the relatively low stellar density across the Magellanic Bridge. It was shown by \cite{Niederhofer2018a}, that PSF and aperture photometries deliver the same results in regions of low stellar density. The astrometric calibration of the VMC data is based on the Two Micron All Sky Survey \citep[2MASS, ][]{Skrutskie2006} and carries a systematic uncertainty of 10$-$20 mas due to World Coordinate System errors\footnote{http://casu.ast.cam.ac.uk/surveys-projects/vista/technical/astrometric-properties}. Those are systematic uncertainties in the calibration of each detector image obtained using 2MASS stars. They are mainly caused by atmospheric turbulence and atmospheric differential refraction. Table~\ref{Table1} provides details about the observations. It contains the tile identification, the central coordinates, the orientation, the number of epochs used, their time baseline, the FWHM, the airmass and the sensitivity\footnote{http://casu.ast.cam.ac.uk/surveys-projects/vista/technical/vista-sensitivity}, derived from sources with photometric uncertainties <0.1 mag. The average values of all good quality deep $K_{s}$ epochs were 0.95$\pm$0.03~arcsec~(FWHM), 1.58$\pm$0.03~(Airmass) and 19.91$\pm0.06$~mag~(sensitivity).

%-----------------------------------------------------------------------
\section{Analysis} \label{Analysis}
\subsection{VMC proper motions} \label{VMCpms}
\subsubsection{Sample selection} \label{VMCsample}
The VMC source catalogues for each tile were obtained from the VSA using a freeform SQL query. We extracted equatorial coordinates (right ascension, declination) in J2000, source-type classifications ($\textit{mergedClass}$), magnitudes ($J$ and $K_{s}$), the corresponding uncertainties and quality extraction flags for each source. VMC tiles, pawprints and sources in the VSA are identified by their identification numbers: tiles by a unique framesetID, individual pawprints by a unique multiframeID and individual sources by a unique sourceID. The source-type classification flags were used to distinguish between stars and galaxies while quality extraction flags were used to remove low-quality detections. The VSA vmcdetection table contains data of the individual pawprints originating from stacked images. Source catalogues based on individual epoch observations were obtained by cross-matching the list of sources with those in the vmcdetection tables (i.e. tables generated from pawprint images) retaining all matches within 0.5$^{\prime\prime}$. The resulting catalogue contains the mean Modified Julian Day (mjd) of the observation, the detector number (extNum), the pixel coordinates ($x$, $y$) on each detector and the corresponding positional uncertainties.

We split the VMC epoch catalogues into 96 parts (for 6 pawprints $\times$ 16 detectors) per epoch and tile, respectively. A post-processing error quality bit flag (ppErrBits), which is a useful flag to remove spurious detections, was set at 16 or smaller. This selection criterion removes VMC sources with systematics affecting the photometric calibration. Distinct epochs were then selected based on their multiframeIDs. All sources with the same multiframeID are part of the same pawprint observed across the 16 detectors. 
Undesired multiframeIDs such as those associated with observations from overlapping tiles (where sources would be detected in different detectors), observations obtained during poor sky conditions, and detections at wavelengths other than $K_{s}$ were removed. Every catalogue was then divided into two parts. One contains only sources classified as galaxies ($\textit{mergedClass}$=$1$) and the other contains only stars ($\textit{mergedClass}$=$-$1). We rejected all other source-type classifications (e.g.~noise, probable stars and probable galaxies). In a small number of cases two sources in the vmcdetection tables, with the same multiframeIDs, were matched to one sourceID from the vmcsource catalogue. This duplication was caused by the matching algorithm when two sources were sufficiently close together in the detection catalogue while one of the sources was missing in the vmcsource catalogue. The nearest source was selected in this case.

\subsubsection{Astrometric reference frame} \label{referenceframe}

To calculate consistent proper motions, each observation of a given source has to be in the same astrometric reference frame. The reference frames for each VMC tile in this study were created by choosing the epoch with the best observing conditions from each set of observations of a given tile. This corresponds to the epoch with the most significant number of extracted sources and the smallest FWHM.
In a pilot study \citep{Cioni2013}, where proper motions were calculated on a tile-by-tile basis, and in a subsequent study \citep{Cioni2016}, where better proper motions were obtained on a detector-by-detector basis, a reference system was constructed by using background galaxies. The number of background galaxies in the VMC survey is quite large (a few hundred per detector) and using them to build reference frames seems reasonable. However, the extended nature of galaxies increases the position uncertainties, so averaging a sufficient number (>100 per detector) of them is necessary for better results. There are too few point-like background sources such as background quasars. Unresolved background galaxies are more numerous than quasars, but a clear selection is very challenging.

In a recent study \citep{Niederhofer2018a}, we created the reference frames by using the more numerous stars of the 47 Tuc star cluster. This method yielded more accurate proper motions, mainly because of smaller rms residual values of the matching \citep[rms<0.09 pixels,][]{Niederhofer2018a}, achieved through better positional matches between the epochs by significantly increasing the number of reference sources per detector. There was less improvement in the rms values for areas containing more than $\sim$400 reference sources per detector. However, using stars as reference sources is less feasible in the sparsely populated Magellanic Bridge, where background galaxies outnumber actual Bridge stars. In this study we used background galaxies for the whole VMC Bridge area, although we could have also used Bridge stars in more populated regions closer to the LMC and SMC to obtain similar results. The median rms values of matching the epochs using background galaxies was 0.24 pixels and all matches had values smaller than 0.3 pixels. These residuals affect the median proper motion of the background galaxies, resulting in a moving reference frame.
To correct the co-moving reference frame, we set the sigma-clipped relative median proper motion of the galaxies for each detector to zero. We checked for possible systematics, such as unevenly distributed samples, an influence of uncertainties in individual coordinates and the size of the matching samples. The size of the matching samples seemed to have the most significant influence on the results. In two tiles where galaxies were significantly outnumbered by stars, using stars to create the reference frame performed better (rms<0.22~pixels) than using galaxies. The results in regions with similar numbers of stars and galaxies were comparable, while slightly favouring the use of stars due to better centroid determinations (two tiles). In regions with significantly fewer stars than galaxies, which is the case in 9 out of 13 tiles centred on the Magellanic Bridge, the reference frame constructed from background galaxies provided significantly better results. We decided to use only background galaxies for consistency since the majority of the tiles contain more galaxies than stars.
%-----------------------------------------------------------------------
\subsubsection{Deriving the proper motions}\label{pm}

After choosing the reference frame, every corresponding epoch catalogue was transformed into it using IRAF tasks {\it xyxymatch}, {\it geomap}, and {\it geoxytran} and then joined with the reference epoch catalogue of the same pointing and detector. The proper motions of individual stars were calculated by using a linear least-squares fit for the $x$ and $y$ coordinates separately, and the corresponding mjd with respect to a reference frame defined by background galaxies. Each fit contained on average 10 data points, with a minimum of 8, spanning an average time baseline of 921 days (see Table~\ref{Table1} for the time baseline for each tile). Calculations were performed on a detector-by-detector level for each of the 16 detectors and 6 pawprints of each tile. The slopes of these fits are the proper motions of individual stars for the two components in units of pixels per day. 
A Bayesian fitting method to obtain the linear fits was also tested, but it did not show any significant improvement, that would justify the increase in calculation time.
The proper motion results correspond to the reference frame, so the proper motion d$\nu$ and d$\eta$ in pixel per day on the detector axes correspond directly to $\mu_{\alpha}\cos{\delta}$ and $\mu_{\delta}$. Following the convention adopted by \citet{Cioni2016}, we converted the proper motions from pixel per day to mas per yr. When calculating the median proper motions of a selection of stars, we removed outliers using a 3$\sigma$ clipping technique where $\sigma$ was calculated using the median absolute deviation (MAD),

%equation
\begin{equation}
    MAD=\mathrm{median}(\left|X_{i}-\overline{X}\right|),
\end{equation}

where $X$ represents the proper motion measurements of a given sample. The statistical error was calculated as the MAD divided by the square root of the numbers of stars. We used the MAD because it is less influenced by outliers than the mean ($\mu$) and standard deviation ($\sigma$). For symmetric Gaussian distributions $\sigma$ and MAD are related through:

%equation
\begin{equation}
    \sigma\approx 1.4826 \times MAD.
\end{equation}

The standard deviation was then used to exclude sources more distant than $3 \sigma$ from the median of a given sample. The sigma clipping was repeated until no additional sources were removed. We used a 2$\sigma$ clipping to further remove outliers to reduce the influence of MW foreground stars. The median proper motion of the background galaxies of the corresponding detectors was then calculated and its values were subtracted from the corresponding stellar proper motions. We checked the proper motions for any trends with detector number, position on the detectors and $J-K_{s}$ colours and found nothing significantly influencing our results. 

\subsection{$Gaia$~DR2 sample selection and comparison to VMC data}\label{Gaiasample}

We acquired $Gaia$~DR2 data through the $Gaia$@AIP database\footnote{https://gaia.aip.de}, to further improve our VMC sample selection and to obtain a comparable catalogue of proper motions. We used the selection criteria recommended by the $Gaia$ data processing and analysis consortium (DPAC) on the $Gaia$~DR2 'known issues' website\footnote{https://www.cosmos.esa.int/web/gaia/dr2-known-issues} to select well-measured sources. Our selection uses the $Gaia$ Astrometric goodness-of-fit flag ($\chi^{2}$) and the number of good observations (N) provided in the $Gaia$~DR2 catalogues, to derive the unit weight error (UWE) using
\begin{equation}
    UWE= \sqrt{\chi^{2}/(N-5)}.
\end{equation}

\begin{figure}
   \centering
   \includegraphics[width=0.49\textwidth]{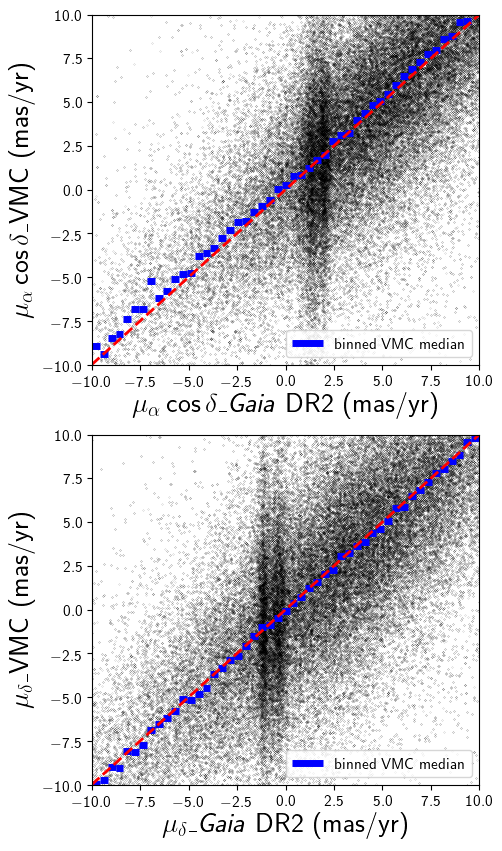}
   
   \caption{VMC proper motions as a function of their corresponding $Gaia$~DR2 proper motions for all stars of the cross-matched catalogues. Regions of highest density correspond to the MCs. They are visible as two vertical features in both panels. The left feature corresponds to the SMC (top: $\sim$1.2~mas~yr$^{-1}$, bottom: $\sim-$1~mas~yr$^{-1}$), while the right feature represents the LMC (top: $\sim$2~mas~yr$^{-1}$, bottom: $\sim-$0.5~mas~yr$^{-1}$). The red dashed line shows the one to one correlation and the median VMC proper motions are shown in blue.}
    \label{PM_compare_GaiaDR2_VMC}% 
\end{figure}

An empirical normalisation factor ($u_{0}$) for the UWE is provided in a lookup table on the ESA $Gaia$~DR2 'known issues' page. This factor is a function of magnitude (G) and colour ($C =G_{BP}-G_{RP}$). The $u_{0}(G, C)$ in our case (mainly for G>15~mag) was $\sim$1 for all sources and therefore we assumed the renormalised unit weight error RUWE ($= UWE/u_{0}$) to be equal to the UWE. We selected only sources with $RUWE< 1.40$ to remove stars that could be problematic sources such as astrometric binaries, (partially) resolved binaries or multiple stars blending together. It was not necessary to remove brighter stars (G<15~mag) or based on their proper motion (e.g. fast-moving foreground stars) at this stage, since subsequent steps removed those efficiently. For all $Gaia$~DR2 proper motions we took the error correlation between proper motion in right ascension and declination into account. The significant correlations are a direct result of the simultaneous five-parameter fit \citep[position, parallax, proper motion; see][]{Luri2018}. On the contrary, VMC proper motion errors in both directions are not correlated, since they are computed separately. We cross-matched the VMC and $Gaia$~DR2 catalogues taking $Gaia$~DR2 proper motions into account by using the J2000 coordinates of the $Gaia$~DR2 sources within a radius of 1$^{\prime\prime}$. We compared the proper motions from VMC and $Gaia$~DR2 to check for systematics. In Fig.~\ref{PM_compare_GaiaDR2_VMC} we show VMC proper motions as a function of their corresponding $Gaia$~DR2 proper motion. VMC proper motions show a large spread and therefore prove to be unreliable on the level of an individual source. However, medians of VMC proper motions resulting from
binning large samples show a good agreement. This holds especially true for the range where we measure the proper motions of the MCs. We did not find any strong correlation on brightness, colour or sky position in that range. Sources outside this range are MW foreground stars and will be removed by subsequent steps in the next section. The comparison does not show all VMC sources, since the VMC data contain a significantly larger number of fainter sources that are not covered by $Gaia$. Covering fainter sources significantly increases the sample size by probing a lower stellar mass regime (see Sect.~\ref{compare}). This makes VMC proper motions a valuable addition. We hope to reduce the spread in the VMC proper motions in the future (see Sect.~\ref{Discussion}).

\subsection{Milky Way foreground simulation: GalMod} \label{GalMod}

\begin{figure*}
   \centering
   \includegraphics[width=1.0\textwidth]{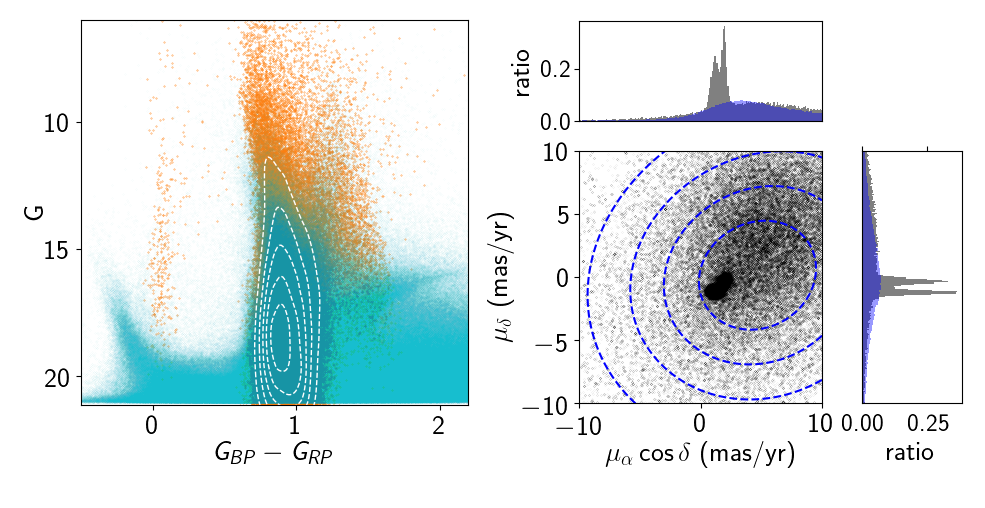}
   
   \caption{Left: CMD of the GalMod theoretical population synthesis model (orange) overlayed on all $Gaia$~DR2 sources (cyan) of the same area of the Magellanic Bridge. The white contour levels show the model number density distribution in steps of 20,000 sources starting at the centre. Right: Proper motion distribution in right ascension and declination of $Gaia$~DR2 sources (black). The contour levels of the foreground simulation are indicated with blue dashed lines and contain each 20,000 sources. The corresponding ratios between the two proper motion components for the $Gaia$~DR2 proper motions are shown in grey and for the GalMod simulation in blue. The LMC and the SMC are visible as two narrow peaks, while the MW foreground stars are more spread out.}
    \label{Figure_GalMod}% 
    \end{figure*}

We used the theoretical population synthesis model GalMod\footnote{https://www.galmod.org/} \citep[][]{Pasetto2018} to investigate the influence of the MW foreground stars in our VMC and $Gaia$~DR2 sample, since it provides photometry and kinematics. GalMod can simulate synthetic surveys of the MW in a selected area of the sky including information about non-axisymmetric features such as spiral arms and bar. Additionally, it includes a geometry-independent ray-tracing extinction model based on the DART-ray radiation transfer code \citep{Natale2017}. Substructures such as satellite galaxies and streams are not included in the model. GalMod data can be queried in a variety of photometric systems (e.g. 2MASS and $Gaia$). We used the online form to select a rectangle containing the Magellanic Bridge, the $Gaia$ photometric system and the maximum possible field of view depth of 50~kpc. All other parameters were left at their default values. For this study, we queried the simulated information about positions, distances, proper motions and magnitudes. The queried GalMod data contain 100,534 sources. They are evenly distributed across the Magellanic Bridge and cover an area that fully includes the VMC Bridge area. 

Fig.~\ref{Figure_GalMod} shows a colour-magnitude diagram (CMD) of the GalMod and $Gaia$~DR2 data of the same area in the $Gaia$ photometric system \citep[G$_{BP}$, $G_{RP}$, and $G$;][]{Evans2018}. It shows that the most prominent vertical feature at $G_{BP}-G_{RP}=0.8$~mag is reproduced by the simulation. Another less populated feature is present at $G_{BP}-G_{RP}=0.1$~mag. Features to the left and right of the model, can be associated with the MCs. They refer to young main sequence stars at $G_{BP}-G_{RP}=-0.3$~mag and to RGB stars at $G_{BP}-G_{RP}=1.1-2.0$~mag. From the distances provided by the model we derived theoretical parallaxes in order to test the parallax selection criterion used in the next section. Additional focus was put on the proper motions of the MW foreground stars, to quantify their potential effect on the proper motion measurements of the Magellanic Bridge. 
GalMod and $Gaia$~DR2 proper motions are compared in Fig.~\ref{Figure_GalMod} (right). The central black overdensities are associated with the MCs (LMC at the top and SMC at the bottom). We expect the proper motion of stars belonging to the Magellanic Bridge to be located between these regions. Black dots further away from the centre are likely MW foreground stars. The proper motions of the MCs are found within the highest number density contour-level, where a large number of MW foreground stars have very similar proper motions. This is also visible in the ratios of the two proper motion components, where the wide spread of the MW foreground stars overlaps with the two peaks of the MCs and contributes unevenly. A selection based on proper motions alone, therefore, always contains MW foreground stars and hence biases the results. The influence also increases in the presence of large proper motion uncertainties because of more overlap. The proper motion in $\mu_{\alpha}\cos{\delta}$ shows a larger amount of foreground stars with larger proper motions. Therefore the median proper motion of samples without efficient foreground removal measures larger proper motions in $\mu_{\alpha}\cos{\delta}$. From the GalMod simulation we can get an idea about the direction in which the proper motion measurements of the Clouds are biased. The MW foreground stars in that area of the sky have, on average, larger proper motions than the MCs. This effect should be more significant in $\mu_{\alpha}\cos{\delta}$ than in $\mu_{\delta}$, since the MW foreground proper motions are on average more than three times larger in $\mu_{\alpha}\cos{\delta}$ than in $\mu_{\delta}$. Comparing the distribution predicted by the GalMod simulation with the distribution of $Gaia$~DR2 MW foreground stars we found a larger number of stars with larger absolute proper motions than suggested by the simulation. In particular, the proper motion in $\mu_{\delta}$ shows a large discrepancy. This discrepancy could be due to selection effects based on distance. A star with a given tangential velocity has a smaller proper motion at a larger distance but is also less likely to be observed, and this would explain a lack of stars with small absolute proper motions. However, this does not explain why the discrepancy in both directions is one-sided towards more positive proper motions. An explanation for it could be the Lutz-Kelker bias \citep{Lutz1973}. The one-sidedness may be due to simple statistics: it is easier to scatter out of the volume than into it, given a homogeneous density distribution of objects. Still, both the median $Gaia$~DR2 and VMC proper motions of the MW foreground stars are in good agreement.

\subsection{Milky Way foreground removal} \label{MWremoval}
\subsubsection{Removal with the VMC and $Gaia$~DR2 (method~1)}\label{method1}

    % Bridge CMDs
   \begin{sidewaysfigure}
   \centering
   \includegraphics[width=1.0\textwidth, ]{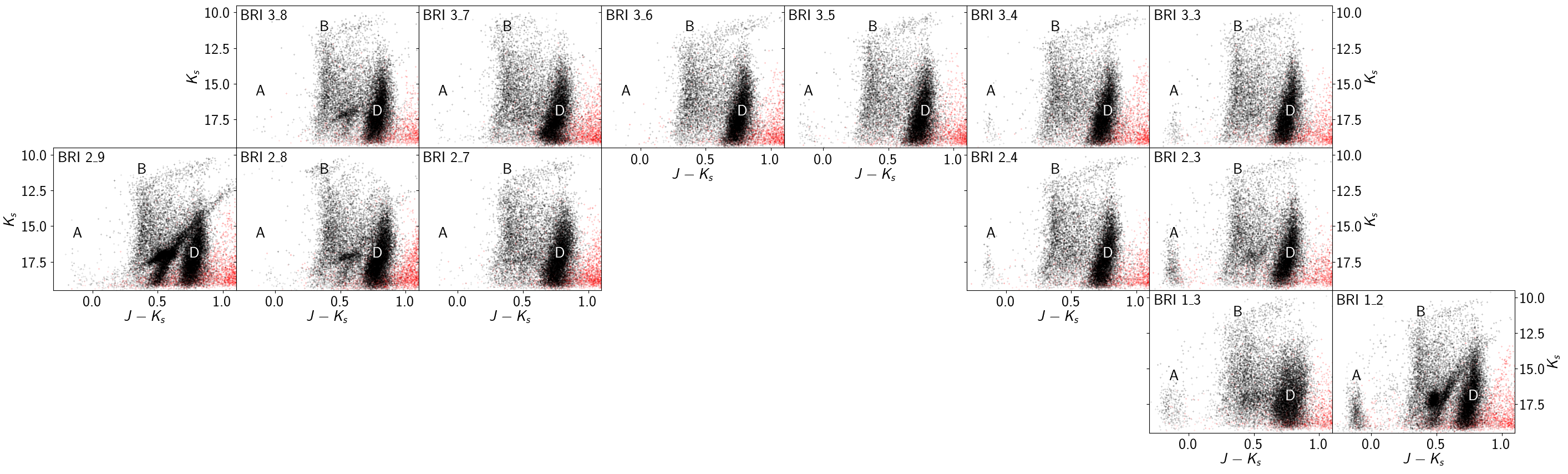}
   \caption{Colour-magnitude diagrams of individual VMC Magellanic Bridge tiles. Stars are shown in black, while sources classified as galaxies are shown in red. Real background galaxies mainly occupy colours $J-K_{s}> 1$~mag and are not shown here to emphasise stellar features, while extended objects with $J-K_{s}<1$~mag are probably the result of stellar blends. Stars associated with the LMC can be seen on the left side (e.g. tile BRI 2$\_$9), while stars on the right, especially in tile BRI 1$\_$2, are associated with the SMC. The prominent features (B) at $J-K_{s}\sim0.4$~mag and (D) at $J-K_{s}\sim0.7$~mag in each panel are caused by MW foreground stars. The majority of stars associated with the MCs can be found at $J-K_{s}\sim0.5$~mag.}
   \label{BRI_13cmds}%
   \end{sidewaysfigure}

    % full cmd
   \begin{figure*}
   \centering
   \includegraphics[width=\textwidth]{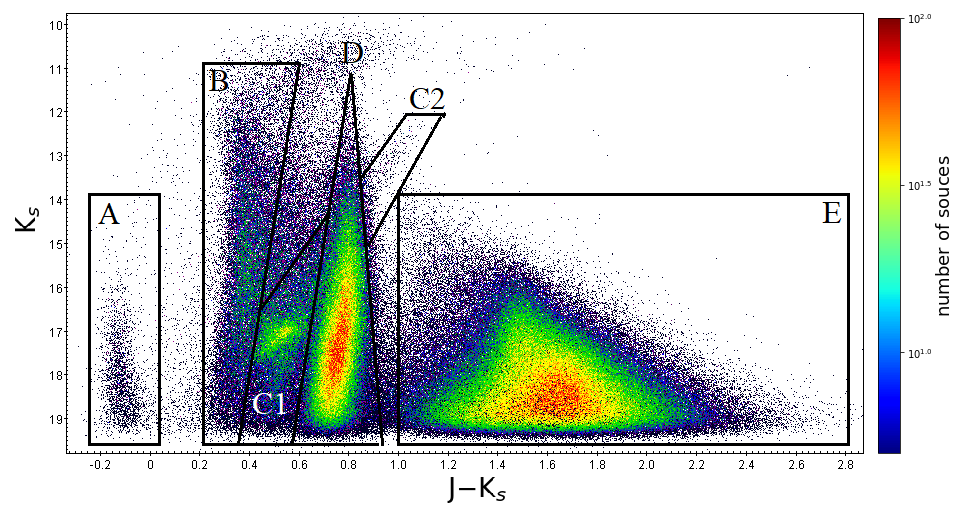}
   \caption{Hess colour-magnitude diagram of VMC sources of the 13 Magellanic Bridge tiles. Regions identified by letters indicate different stellar populations. The majority of stars associated to the MCs can be found at $J-K_{s}\sim-$0.1~mag (A) and $J-K_{s}\sim$0.5~mag.}
   \label{BRI_full_cmd}%
   \end{figure*}

Stars associated with the Magellanic Bridge are more sparsely distributed at the centre of the Magellanic Bridge than towards the LMC and SMC. Fig.~\ref{BRI_13cmds} shows CMDs, ($K_{s}$, $J-K_{s}$), of all 13 VMC-Bridge tiles. Stellar populations within the outskirts of the SMC and from the outer regions of the LMC show clear red clump (RC) and RGB features. These features fade towards the centre of the Magellanic Bridge. The remaining features (B and D) seen in the centre of the Magellanic Bridge (e.g. in tile BRI~3\_6) are related to MW foreground stars and they are visible in all tiles across the Magellanic Bridge. The MW foreground population traces two nearly vertical features at $J-K_{s}=0.35$~mag and $J-K_{s}=0.7$~mag. The main sequence turn-off of various MW populations causes the first feature while the second is caused by the 'CMD kink' of low-mass cool M dwarfs \citep{Rubele2018}.
A Hess colour--magnitude diagram combining all VMC-Bridge tiles is shown in Fig.~\ref{BRI_full_cmd}, where we divide the CMD into regions containing different stellar populations. Region A contains a young stellar population mainly found towards the SMC and believed to have been formed in situ \citep{Irwin1985}. Regions B and D contain mainly MW foreground stars. Region D intersects with the RGB of the MCs and divides it into regions C1 and C2. Region C1 contains the RC and the majority of MC stars and region C2 the tip of the RGB. The connecting region between C1 and C2 is dominated by the MW foreground stars of region D. Region E contains a large number of background galaxies. It is important to note that stars with $K_{s}>19$~mag were removed from our analysis because VMC proper motions become unreliable at this magnitude.
To reduce the influence of MW foreground stars on our samples, we used data from $Gaia$~DR2 cross-matched with data from the VMC survey. We removed stars with absolute parallaxes (from $Gaia$~DR2) larger than 0.2 mas. The VMC$-Gaia$ DR2 cross-matched catalogue contains 179,049 unique sources of which 45,754 have $Gaia$~DR2 parallaxes and only 12,014 have parallaxes smaller than 0.2~mas, that is, they are likely members of the Magellanic Bridge. The bottom panels of  Fig.~\ref{CMD_pcut_SH_30kpc} show the resulting selection. The bottom left panel shows stars that are removed as MW foreground stars by this parallax selection criterion, while the other panels show the remaining stars in the CMDs as observed with the VMC (bottom-middle) and $Gaia$~DR2 (bottom-right). All panels are colour coded by their estimated distance (described in next the section) to allow us to compare both methods. This selection criterion mainly removes foreground stars up to $12$ kpc with a declining efficiency at large distances due to the nature of $Gaia$ parallaxes \citep{Luri2018}. These stars are mainly brighter than $K_{s}\sim$16 mag, except for some fainter blue stars with $K_{s}\sim$18 mag. This selection efficiently removes most of the MW foreground stars of region D. Because of the difficulty of measuring stellar proper motions at the distance of the Magellanic Bridge, $Gaia$~DR2 proper motion errors are too large to separate individual stars in proper motion space allowing for a similar efficient removal. Furthermore, the Magellanic Bridge stellar population density in the central regions is much lower than that of the MW foreground stars. Therefore a combination of CMD selection criteria (Boxes A, C1 and C2; Fig.~\ref{BRI_full_cmd}) and $Gaia$~DR2 parallaxes was necessary to increase the ratio between Bridge and MW foreground stars, without removing many potential Bridge stars. However, in Section~\ref{compare} we show that this selection removes potential Bridge stars while retaining some MW foreground stars.

\subsubsection{Removal with the StarHorse (method~2)} \label{method2}

\begin{figure*}
   \centering
   \includegraphics[width=\textwidth]{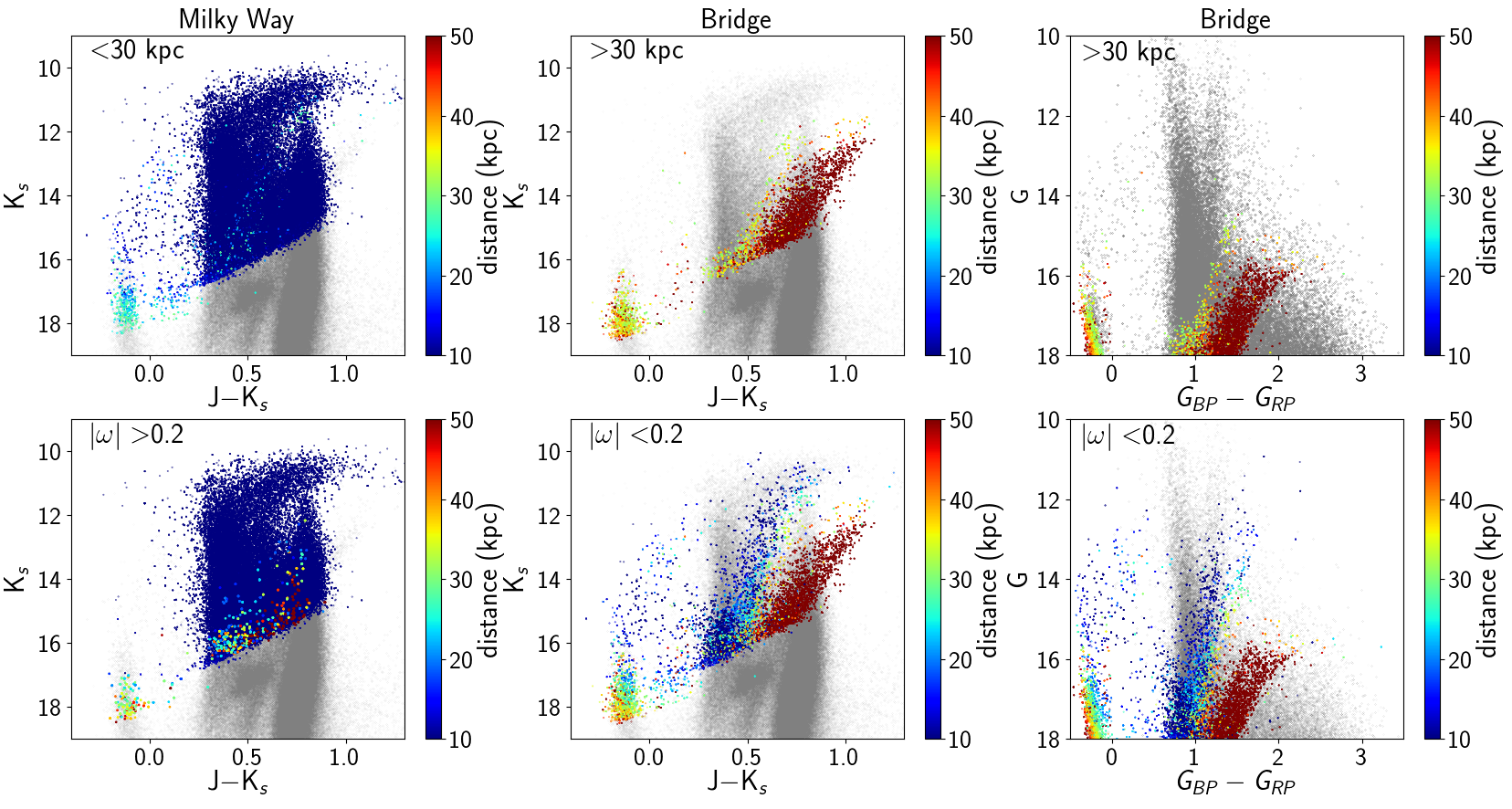}
   \caption{Colour magnitude diagrams of stars within the VMC-Bridge tiles (grey) colour coded by StarHorse distances. Stars are selected based on StarHorse distance (top) and $Gaia$~DR2 parallax (bottom). Bridge stars are shown in the near-infrared CMD (middle) and in the $Gaia$ CMD (right) while Milky Way foreground stars are shown in the near-infrared CMD (left).}
              \label{CMD_pcut_SH_30kpc}%
    \end{figure*}

\begin{figure*}
   \centering
   \includegraphics[width=1.0\textwidth]{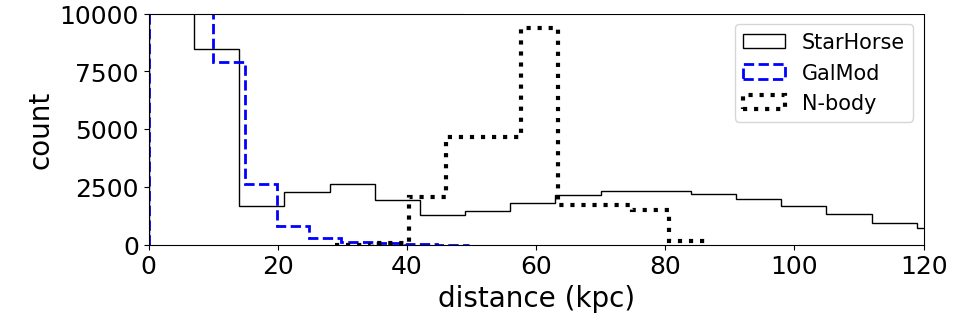}
   
   \caption{Histograms of distances in the Bridge area. The blue dashed line indicates the distribution of the MW foreground stars (GalMod), the black dotted line the distribution of the SMC particles from an $N$-body simulation \citep{Diaz2012} and the solid line the distance estimates of MC stars from StarHorse \citep{Queiroz2018}.}
    \label{histogram}% 
    \end{figure*}

To more efficiently remove MW stars from both our VMC and $Gaia$~DR2 crossmatched sample of Magellanic Bridge stars we also probed the benefits of a more sophisticated tool to analyse astrometric and photometric data, StarHorse \citep{Queiroz2018}. This is a Bayesian tool for determining stellar masses, ages, distances, and extinction values of field stars. It is based on a Bayesian inference code first presented by \citet{Santiago2016}. In this study, we used distances derived with this code based on $Gaia$~DR2 data for sources with $G<18$~mag. This limitation was imposed by the Bayesian tools being computationally heavy and the challenging numbers of $Gaia$~DR2 sources. Applying the code to fainter sources is possible, although less reliable, due to the increase of errors, but we plan to investigate this further in the future. From this point on, we refer to this sub-sample of the $Gaia$~DR2 data whenever we mention distances as the StarHorse sample. This sub-sample covers only the upper RGB of the MCs and stops right above the RC (see middle panel of Fig.~\ref{CMD_pcut_SH_30kpc}). The top panels of Fig.~\ref{CMD_pcut_SH_30kpc} show a more efficient removal of MW foreground stars compared with the previous selection (top panel). However, it should be clearly stated that the purpose of StarHorse is to provide distance estimates for MW stars using priors that reflect the properties of the MW. Therefore stars associated with the MCs are not expected to have precise distance estimates. However, due to their $Gaia$~DR2 parameters (e.g. parallaxes, G magnitude and B$_{BP}-$B$_{RP}$) they are expected to end up at large distances.  A histogram of the distance distribution is shown in Fig.~\ref{histogram}.

 The distances estimated by StarHorse are compared with the distance distribution from the GalMod simulation. Both distributions agree up to around 20~kpc. Then, the number of stars in the StarHorse sample starts to increase. This behaviour would not be expected for the MW. Therefore StarHorse suggests that an increase is more likely associated with the MCs. We also produced a histogram of the distances of stripped SMC particles from an $N$-body simulation by Diaz \& Bekki (2012; see Section~\ref{nbody}). Those particles, however, represent mass particles and not individual stars. Individual stars would show a larger spread.
According to the \citet{Diaz2012} simulation, the Bridge stars should be found at a distance between 40 and 80 kpc, which is in agreement with our current understanding that the Magellanic Bridge stretches mainly along the line of sight between the LMC and SMC \citep[e.g.][]{JacyszynDobrzeniecka2020}. Most Magellanic Bridge stars should have distances of 50$-$60~kpc. StarHorse distances for the Magellanic Bridge do not indicate a peak in the distribution at that point. On the contrary, there is a dip in the distance distribution around 50~kpc. We assume this to be caused by using MW priors for the MCs. We found that the expected stellar population at distances of 50$-$60~kpc seems to split and shift. There are two peaks in the distribution around 30 and 80 kpc. We found that stars at 30~kpc were on average bluer in $G_{BP}-G_{RP}$ colours compared with stars at 80 kpc. In summary, we used a StarHorse distance of more than 30~kpc to select Magellanic Bridge stars (see Fig.~\ref{histogram}).

\subsection{$N$-body simulations of the Magellanic Bridge} \label{nbody}

According to recent studies \citep[e.g.][]{Zivick2019} there is no reason to doubt that the Magellanic Bridge was formed by the last interaction between the LMC and SMC. During that interaction the tidal forces affecting the SMC can be assumed to have been significantly larger than those affecting the LMC, due to the large difference in mass between the LMC and SMC. However, such an interaction is very complex and depends on many parameters other than only mass. The involvement of the MW further complicates matters by introducing a three-body-problem. Many studies \citep[e.g.][]{Murai1980,Gardiner1996} have suggested that the Magellanic Bridge consists mainly of material stripped from the SMC, which is supported by many findings \citep[e.g.][]{Irwin1985, Dufton2008, Carrera2017} and recently by \citet{DeLeo2020}, where they suggest a net outward motion of stars from the SMC centre along the direction towards the LMC. Therefore we compared our measurements with an $N$-body simulation introduced by \citet{Diaz2012}. In this simulation the SMC is represented by three components: disc, spheroid and dark matter halo. The simulation describes the tidal evolution of both the disc and the spheroid component of the SMC based on HST proper motions \citep{Kallivayalil2006a, Kallivayalil2006b}. In that simulation the SMC ($m_{\text{SMC}}=3\times 10^{9} \textit{M}_{\odot}$) interacts with the LMC ($m_{\text{LMC}}= 10^{10} \textit{M}_{\odot}$), the latter in the form of a static potential (including a dark matter halo), over a period of more than 3 Gyr. Arriving at the current position of the SMC, the simulation reproduces the Magellanic Stream and Bridge. We used current day kinematics from the simulation (positions and velocities) to compare with our proper motion measurements. The model is not consistent with the latest improvement of the $Gaia$~DR2 measurements of the MCs (e.g. LMC and SMC proper motions), but HST measurements \citep[][]{Zivick2019} show similar proper motions for the Magellanic Bridge (see the discussion in Section~\ref{Discussion}). Most simulations try to reproduce the location of the Magellanic Stream \citep[e.g.][]{Diaz2012} and the $Gaia$~DR2 proper motions of the Clouds, to constrain their models. Like the Magellanic Stream the Magellanic Bridge is also a relic of past interactions. Therefore the proper motion of the Magellanic Bridge can be used as an additional independent constraint on the MC models aiming at describing the more recent evolutionary history of the galaxies.

\section{Results} \label{results}

\subsection{Comparing the Milky Way removal methods} \label{compare}
    
In the previous section, we discussed the removal of a significant fraction of MW foreground stars ($>85\%$) by a simple selection in absolute parallax ($|\omega|>0.2$~mas) combined with a CMD selection based on the CMD regions C1 and C2 (Fig.~\ref{BRI_full_cmd}). Then, we introduced a method based on estimated StarHorse distances. In Fig.~\ref{CMD_pcut_SH_30kpc} we compare the two methods. Both methods agree well in flagging MW foreground stars in most cases. There are however sources that should belong to the MCs according to their StarHorse distance estimation but that are removed by the parallax selection, while other sources apparently belonging to the MW foreground are associated with the MCs. Those sources occupy regions of the CMD typical of MW foreground stars (see Fig.~\ref{BRI_full_cmd}). 

The parallax selection combined with a CMD selection provides a suitable option to remove MW foreground stars. However, compared to the StarHorse method, we found that this first method is less efficient in the central regions of the Magellanic Bridge, which show a lower density of Magellanic Cloud stars. 
In Fig.~\ref{PM_map_method1} we show that the median proper motion towards the central regions of the Magellanic Bridge is influenced by the median proper motion of stars belonging to the MW foreground in both methods. This effect is directly related to the influence of MW foreground stars in the samples drawn from a sparsely populated region. A multitude of factors cause this difference. There is a bias towards bright stars. The most luminous and numerous stars observed in the MCs are RGB stars. The main sequence is often only represented by the most luminous blue stars around the turn-off, which represent the massive stars of a given stellar population. The more numerous low mass main sequence stars are too faint for the VMC and $Gaia$ surveys. Low mass MW stars therefore quickly outnumber the more distant stellar populations when covering large areas of the sky since their distances are smaller. Magnitudes and colours of MW stars can be similar to those of RGB stars at large distances. 
Precise distance measurements of faint stars proved to be challenging. Therefore, kinematics are often used to discriminate between stellar populations (e.g. streams and open clusters). Stars of the same host tend to display similar kinematics. This similarity may be visible as clustering of their proper motions unless proper motion uncertainties and intrinsic spreads within the stellar populations dominate. Discriminating between two populations can be challenging, when either their kinematics are too similar or they overlap. Indeed this is the case for the MW foreground stars and the stellar population of the Magellanic Bridge (see Fig.~\ref{Figure_GalMod} right). This is not an issue in dense stellar regions, where stars of a given stellar population outnumber the MW foreground stars. Hence, we did not select stars solely on the base of proper motions. The final catalogue resulting from the first method contains 14,725 unique sources that satisfy the CMD selection (Fig.~\ref{BRI_full_cmd}) and were not flagged as foreground stars based on their parallaxes. The final catalogue resulting from the second method, by selecting sources based on their StarHorse distance contains instead 3326 sources. 

\subsection{Proper motion of the Magellanic Bridge} \label{Bridgepm}
\begin{figure*}
   \centering
   \includegraphics[width=\textwidth]{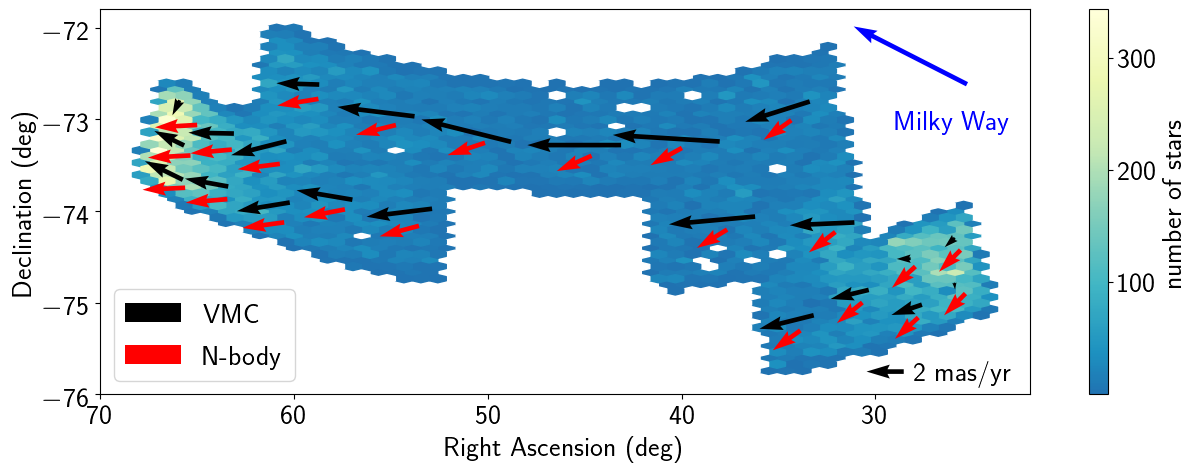}
   \includegraphics[width=\textwidth]{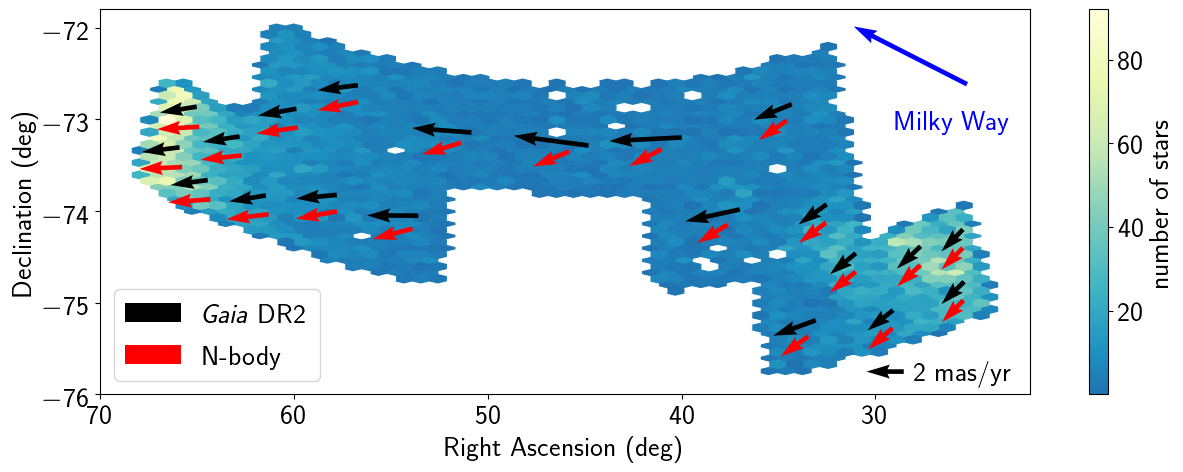}
   \caption{Proper motion maps of the Magellanic Bridge based on method~1 from VMC (top) and $Gaia$~DR2 (bottom) compared with the proper motions obtained from a dynamical model of the SMC$-$LMC interaction (red arrows). The median proper motion of foreground Milky Way stars (top: $|\omega|>0.2$~mas; bottom:$<30$~kpc) is indicated (blue arrows). The background images show the corresponding stellar densities for VMC (top) and $Gaia$~DR2 (bottom) sources.}
              \label{PM_map_method1}%
    \end{figure*}

We used two-dimensional Voronoi binning\footnote{https://pypi.org/project/vorbin/} \citep{Cappellari2003} to divide the Magellanic Bridge into spatial bins each containing a minimum of 250 stars for the first method and 25 stars for the cleaner sample (second method). Both VMC and $Gaia$~DR2 proper motions were compared to a dynamical simulation where the SMC experienced significant stripping as a consequence of its interaction with the LMC (\citealt[][]{Diaz2012}; see Section~Sect.~\ref{nbody}). The two-dimensional Voronoi binning code was applied to the simulated particles together with the stars to avoid truncation effects caused by the edges of the VMC tiles. Simulated particles were given 'no signal' in Voronoi binning so as not to influence the binning of the stars. This binning enables direct comparisons between the model and the two proper motion catalogues.
The median proper motion of a MW foreground sample was calculated by selecting stars in regions B and D (see~Fig.~\ref{BRI_full_cmd}). We found that the average median proper motion of the MW foreground stars is consistent for both catalogues in the central regions of the Magellanic Bridge ($6.09\pm0.01$ mas~yr$^{-1}$ in right ascension and $3.10\pm0.01$ mas~yr$^{-1}$ in declination), but not with the GalMod simulation, which suggests a smaller
proper motion in declination ($\sim$0.5~mas~yr$^{-1}$~see right panel Fig.~\ref{Figure_GalMod}) perhaps due to a selection effect based on distance. Fig.~\ref{PM_map_method1} shows the resulting proper motion maps from VMC (top) and $Gaia$~DR2 data (bottom) using the first method to remove MW foreground stars. Both data sets indicate that stars move from the SMC towards the LMC. This motion was first shown by \citet[][]{Schmidt2018} and confirmed by \cite{Zivick2019}. The two maps display a similar trend in the central region of the Magellanic Bridge. Proper motions become larger with decreasing stellar density, mainly along right ascension, but also in declination. The VMC proper motions are strongly affected by this and exhibit overall less ordered motions in comparison with $Gaia$~DR2 proper motions. In regions of high stellar density, there is also a good agreement between $Gaia$~DR2 proper motions and the $N$-body simulation of \cite{Diaz2012}. This agreement is not always present with respect to the VMC proper motions. The median proper motions for method~1 shown in Fig.~\ref{PM_map_method1} are summarised in Table~\ref{TablePMVMC} (VMC) and Table~\ref{TablePMGDR2} ($Gaia$~DR2).

\begin{figure*}
   \centering
   \includegraphics[width=\textwidth]{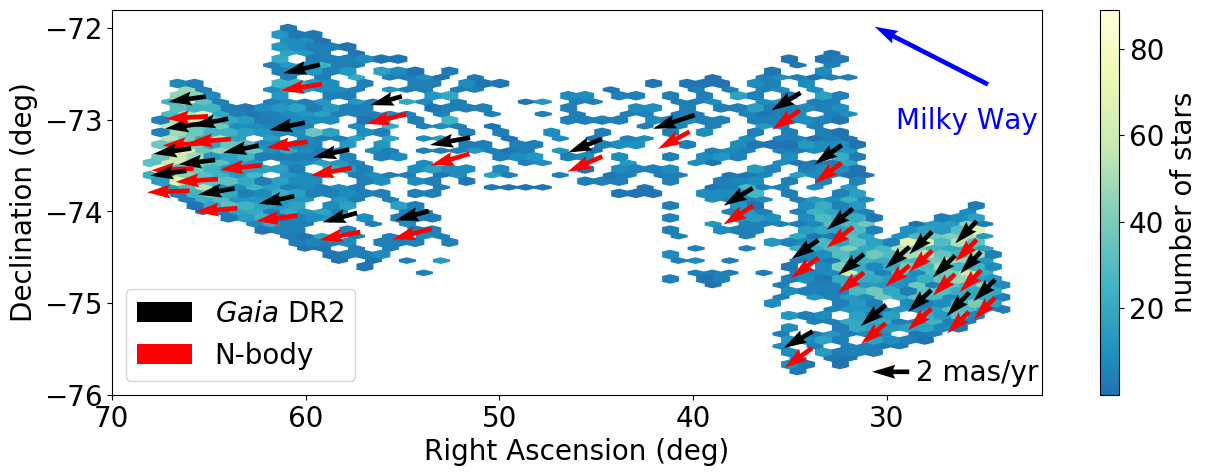}
   \includegraphics[width=\textwidth]{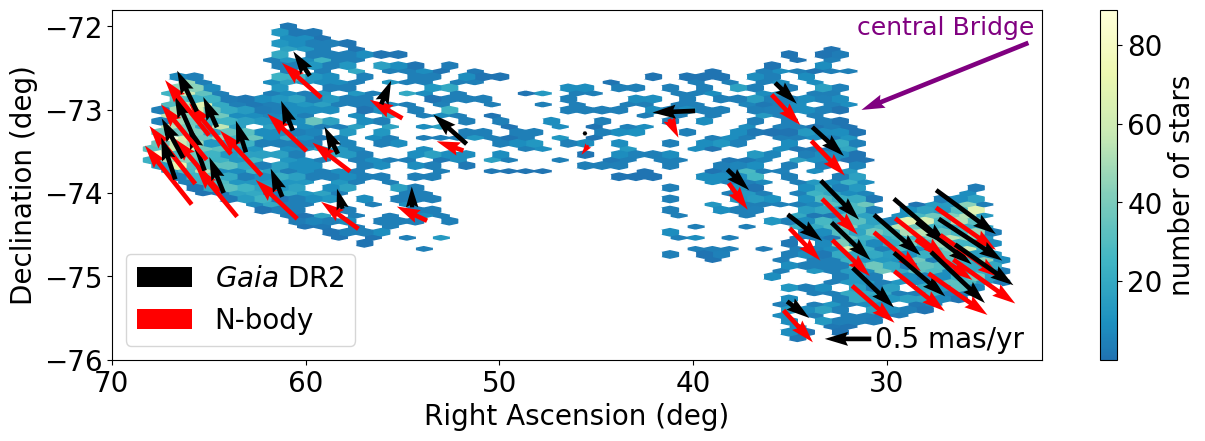}
   \caption{Proper motion map (top) and residual proper motion map (bottom) of the Magellanic Bridge using $Gaia$~DR2 proper motions for bins of stars with StarHorse distances $>30$~kpc (black arrows). The median proper motion of the foreground Milky Way stars (with distances $<30$~kpc) and the median proper motion of the Magellanic Bridge are indicated (blue and purple arrows, respectively).}
              \label{PM_map_SH}%
    \end{figure*}

The resulting VMC and $Gaia$~DR2 proper motion maps of the sample selected using StarHorse distances are shown in Fig.~\ref{PM_map_SH} (VMC at the top and $Gaia$~DR2 at the bottom). Compared with the two previous maps, the effect of large proper motions in the central regions is significantly reduced. This means a reduced influence of MW foreground stars. The median $Gaia$~DR2 proper motions are very similar to the \cite{Diaz2012} simulation. The median proper motion of 63 sources in the bin (\#17) closest to the centre is $1.80\pm0.23$ mas~yr$^{-1}$ in right ascension and $-0.72\pm0.13$ mas~yr$^{-1}$ in declination. The median $Gaia$~DR2 proper motion values from the StarHorse-based sample are summarised in Table~\ref{TableSH}. The neighbouring bins \#16 and \#18 show slightly larger proper motions. Both bins exhibit similar stellar densities. We expect fewer actual Bridge stars in bins \#16 and \#18 as suggested by the distribution of \ion{H}{I} gas, which shows a high density in the centre and two less dense regions on either side (see~Fig.~\ref{Figure1}). Combining bin \#17 and bin \#18 (neighbouring the LMC) does not significantly change the result, while including bin \#16 (neighbouring the SMC) leads to a larger standard error and values closer to the expected proper motion of the MW foreground. This suggests the presence of a higher fraction of MW foreground stars in bin \#16. We created a residual map by subtracting the proper motion of the central bin from the other bins (see bottom panel of  Fig.~\ref{PM_map_SH}). This new map shows that the Bridge is stretching, it is supported by a good agreement between the measurements and $N$-body simulation on the SMC side, while there is less agreement on the LMC side. However, this could be caused by the internal kinematics  (rotation pattern) of the LMC which is not part of the simulation. The higher fraction of MW foreground stars in bin  \#16 is more pronounced. There are not enough stars in the VMC-StarHorse catalogue to provide a reliable VMC proper motion of the Bridge centre using the cleanest sample. It shows only small improvements over the method in some regions and is included here for completeness.
    
    %TablePMVMC
    \begin{table}[]
    \caption{VMC Proper motions of the Magellanic Bridge resulting from a CMD and parallax selection criteria.}
    \label{TablePMVMC}
    \begin{tabular}{cccccc}
    \hline\hline
    Bin &  d\tablefootmark{*}& stars   & $\mu_{\alpha}\cos{\delta}$     & $\mu_{\delta}$&$\cos{}\alpha$\tablefootmark{**}    \\
        &  (deg)                     & & (mas~yr$^{-1}$) & (mas~yr$^{-1}$)& \\
    \hline
    1 & 12.33 & 458 & 4.15$\pm$0.59 & 0.51$\pm$0.58 &  0.936 \\
    2 & 13.53 & 687 & 2.30$\pm$0.28 & 0.07$\pm$0.29 &  0.981 \\
    3 & 12.76 & 542 & 3.00$\pm$0.75 & 0.51$\pm$0.74 &  0.937 \\
    4 & 11.68 & 481 & 3.52$\pm$0.41 & -0.43$\pm$0.42 &  0.990 \\
    5 & 10.98 & 418 & 4.82$\pm$0.59 & 1.20$\pm$0.64 &  0.845 \\
    6 & 9.40 & 391 & 5.04$\pm$0.65 & 0.00$\pm$0.64 &  0.915 \\
    7 & 13.56 & 473 & 2.84$\pm$0.89 & -0.46$\pm$0.95 &  1.000 \\
    8 & 13.87 & 535 & 2.97$\pm$0.69 & -0.75$\pm$0.80 &  0.993 \\
    9 & 14.56 & 591 & 2.45$\pm$0.39 & 0.05$\pm$0.35 &  0.994 \\
    10 & 14.40 & 834 & 2.33$\pm$0.32 & 0.42$\pm$0.34 &  0.966 \\
    11 & 15.03 & 910 & 2.04$\pm$0.30 & 1.00$\pm$0.31 &  0.876 \\
    12 & 15.10 & 895 & 1.60$\pm$0.28 & 0.78$\pm$0.30 &  0.875 \\
    13 & 15.15 & 900 & 0.39$\pm$0.32 & -0.81$\pm$0.33 &  0.489 \\
    14 & 8.07 & 365 & 5.75$\pm$0.68 & 0.35$\pm$0.71 &  0.846 \\
    15 & 6.43 & 600 & 3.47$\pm$0.49 & -1.09$\pm$0.42 &  0.952 \\
    16 & 7.31 & 573 & 4.63$\pm$0.50 & -0.42$\pm$0.49 &  0.897 \\
    17 & 6.35 & 275 & 2.92$\pm$0.65 & -0.73$\pm$0.76 &  0.929 \\
    18 & 5.74 & 1117 & 3.47$\pm$0.55 & -0.15$\pm$0.59 &  0.818 \\
    19 & 5.50 & 768 & 2.05$\pm$0.44 & -0.49$\pm$0.47 &  0.902 \\
    20 & 4.66 & 992 & 0.75$\pm$0.34 & -0.01$\pm$0.33 &  0.752 \\
    21 & 4.80 & 424 & 1.64$\pm$0.40 & -0.56$\pm$0.35 &  0.915 \\
    22 & 4.01 & 838 & 0.54$\pm$0.30 & -0.50$\pm$0.30 &  0.999 \\
    23 & 4.10 & 973 & -0.04$\pm$0.25 & -0.37$\pm$0.27 &  0.642 \\
    \hline
    \end{tabular}\\
    \tablefoottext{*}{Angular distance to the SMC centre.}\\
    \tablefoottext{**}{Cosine of the angle between observation and $N$-body simulation proper motion vectors.}
    \end{table}
    
    %TablePMGDR2
    \begin{table}[]
    \caption{$Gaia$~DR2 proper motions of the Magellanic Bridge resulting from CMD and parallax selection criteria.}
    \label{TablePMGDR2}
    \begin{tabular}{cccccc}
    \hline\hline
    Bin &  d\tablefootmark{*}& stars   & $\mu_{\alpha}\cos{\delta}$     & $\mu_{\delta}$ &$\cos{}\alpha$\tablefootmark{**}  \\
        &  (deg)                     & & (mas~yr$^{-1}$) & (mas~yr$^{-1}$)&  \\
    \hline
    1 & 12.93 & 621 & 2.16$\pm$0.10 & -0.27$\pm$0.14 &  0.997 \\
    2 & 13.77 & 595 & 2.06$\pm$0.07 & -0.36$\pm$0.10 &  1.000 \\
    3 & 12.86 & 595 & 2.19$\pm$0.09 & -0.20$\pm$0.12 &  0.996 \\
    4 & 11.79 & 499 & 2.75$\pm$0.14 & 0.02$\pm$0.16 &  0.967 \\
    5 & 11.32 & 497 & 3.19$\pm$0.17 & 0.24$\pm$0.18 &  0.934 \\
    6 & 9.71 & 368 & 4.02$\pm$0.20 & 0.53$\pm$0.23 &  0.866 \\
    7 & 13.80 & 602 & 1.97$\pm$0.05 & -0.33$\pm$0.07 &  0.999 \\
    8 & 15.10 & 1210 & 1.98$\pm$0.02 & -0.31$\pm$0.02 &  0.995 \\
    9 & 14.38 & 671 & 1.97$\pm$0.04 & -0.30$\pm$0.06 &  0.999 \\
    10 & 14.67 & 1050 & 1.99$\pm$0.02 & -0.27$\pm$0.02 &  0.998 \\
    11 & 15.17 & 1148 & 2.01$\pm$0.02 & -0.25$\pm$0.02 &  0.997 \\
    12 & 8.34 & 395 & 3.90$\pm$0.22 & -0.20$\pm$0.20 &  0.908 \\
    13 & 6.52 & 609 & 2.04$\pm$0.14 & -0.83$\pm$0.12 &  0.975 \\
    14 & 7.31 & 573 & 2.93$\pm$0.15 & -0.64$\pm$0.14 &  0.945 \\
    15 & 6.25 & 337 & 2.29$\pm$0.20 & -0.83$\pm$0.19 &  0.960 \\
    16 & 5.87 & 740 & 1.51$\pm$0.09 & -1.08$\pm$0.07 &  1.000 \\
    17 & 5.51 & 797 & 1.40$\pm$0.02 & -1.14$\pm$0.02 &  1.000 \\
    18 & 5.14 & 509 & 1.38$\pm$0.08 & -1.10$\pm$0.06 &  0.999 \\
    19 & 4.60 & 1120 & 1.27$\pm$0.02 & -1.21$\pm$0.01 &  1.000 \\
    20 & 3.97 & 752 & 1.16$\pm$0.02 & -1.18$\pm$0.02 &  1.000 \\
    21 & 4.14 & 1037 & 1.22$\pm$0.02 & -1.20$\pm$0.01 &  1.000 \\
    \hline
    \end{tabular}\\
    \tablefoottext{*}{Angular distance to the SMC centre.}\\
    \tablefoottext{**}{Cosine of the angle between observation and $N$-body simulation proper motion vectors.}
    \end{table}
    
    %TableSH
    \begin{table}[]
    \caption{$Gaia$~DR2 proper motions of the Magellanic Bridge resulting from a selection criterion based on StarHorse distances.}
    \label{TableSH}
    \begin{tabular}{cccccc}
    \hline\hline
    Bin &  d\tablefootmark{*}& stars   & $\mu_{\alpha}\cos{\delta}$     & $\mu_{\delta}$ &$\cos{}\alpha$\tablefootmark{**}   \\
        &  (deg)                     & & (mas~yr$^{-1}$) & (mas~yr$^{-1}$) \\
    \hline
    1 & 13.70 &  97 & 1.97$\pm$0.08 & -0.46$\pm$0.16 &  0.997 \\
    2 & 12.99 &  68 & 1.94$\pm$0.18 & -0.43$\pm$0.33 &  0.999 \\
    3 & 13.69 &  80 & 1.92$\pm$0.03 & -0.40$\pm$0.16 &  0.998 \\
    4 & 12.64 &  49 & 1.86$\pm$0.05 & -0.51$\pm$0.23 &  0.997 \\
    5 & 13.59 & 125 & 1.92$\pm$0.03 & -0.42$\pm$0.02 &  0.997 \\
    6 & 11.67 &  43 & 1.80$\pm$0.27 & -0.50$\pm$0.46 &  1.000 \\
    7 & 14.24 &  97 & 1.90$\pm$0.02 & -0.38$\pm$0.02 &  0.996 \\
    8 & 14.76 &  86 & 1.94$\pm$0.02 & -0.40$\pm$0.02 &  0.992 \\
    9 & 14.76 &  91 & 1.96$\pm$0.02 & -0.24$\pm$0.02 &  0.999 \\
    10 & 14.39 & 115 & 1.95$\pm$0.02 & -0.33$\pm$0.02 &  0.996 \\
    11 & 15.12 & 103 & 2.00$\pm$0.02 & -0.27$\pm$0.02 &  0.997 \\
    12 & 15.17 &  96 & 2.02$\pm$0.02 & -0.30$\pm$0.02 &  0.995 \\
    13 & 15.19 & 118 & 2.01$\pm$0.02 & -0.28$\pm$0.01 &  0.996 \\
    14 & 15.12 & 132 & 1.95$\pm$0.02 & -0.27$\pm$0.02 &  0.995 \\
    15 & 12.30 &  67 & 1.69$\pm$0.31 & -0.47$\pm$0.42 &  1.000 \\
    16 & 11.36 &  85 & 2.15$\pm$0.18 & -0.41$\pm$0.43 &  0.995 \\
    17 & 9.47 &  63 & 1.80$\pm$0.25 & -0.72$\pm$0.13 &  0.999 \\
    18 & 8.13 &  51 & 2.25$\pm$0.33 & -0.74$\pm$0.28 &  0.984 \\
    19 & 7.08 & 140 & 1.56$\pm$0.07 & -0.94$\pm$0.04 &  1.000 \\
    20 & 6.54 &  89 & 1.56$\pm$0.50 & -0.95$\pm$0.33 &  0.998 \\
    21 & 6.15 &  67 & 1.42$\pm$0.06 & -1.01$\pm$0.04 &  1.000 \\
    22 & 5.83 &  83 & 1.46$\pm$0.07 & -1.03$\pm$0.17 &  0.999 \\
    23 & 6.44 &  41 & 1.56$\pm$0.25 & -0.90$\pm$0.24 &  0.995 \\
    24 & 5.64 & 175 & 1.38$\pm$0.02 & -1.14$\pm$0.02 &  1.000 \\
    25 & 5.54 & 246 & 1.38$\pm$0.02 & -1.12$\pm$0.01 &  1.000 \\
    26 & 5.40 & 114 & 1.35$\pm$0.02 & -1.15$\pm$0.02 &  1.000 \\
    27 & 4.92 & 118 & 1.29$\pm$0.03 & -1.16$\pm$0.02 &  1.000 \\
    28 & 4.75 &  89 & 1.24$\pm$0.03 & -1.19$\pm$0.02 &  1.000 \\
    29 & 4.57 & 150 & 1.22$\pm$0.02 & -1.22$\pm$0.02 &  0.999 \\
    30 & 4.32 & 112 & 1.20$\pm$0.03 & -1.18$\pm$0.02 &  1.000 \\
    31 & 3.91 & 114 & 1.15$\pm$0.02 & -1.19$\pm$0.02 &  1.000 \\
    32 & 4.27 &  82 & 1.22$\pm$0.03 & -1.27$\pm$0.02 &  1.000 \\
    33 & 3.95 &  76 & 1.11$\pm$0.03 & -1.17$\pm$0.02 &  1.000 \\
    34 & 3.90 &  64 & 1.17$\pm$0.03 & -1.16$\pm$0.03 &  0.999 \\
    \hline
    \end{tabular}\\
    \tablefoottext{*}{Angular distance to the SMC centre.}\\
    \tablefoottext{**}{Cosine of the angle between observation and $N$-body simulation proper motion vectors.}
    \end{table}

\section{Discussion} \label{Discussion}
\begin{figure*}
   \centering
   \includegraphics[width=\textwidth]{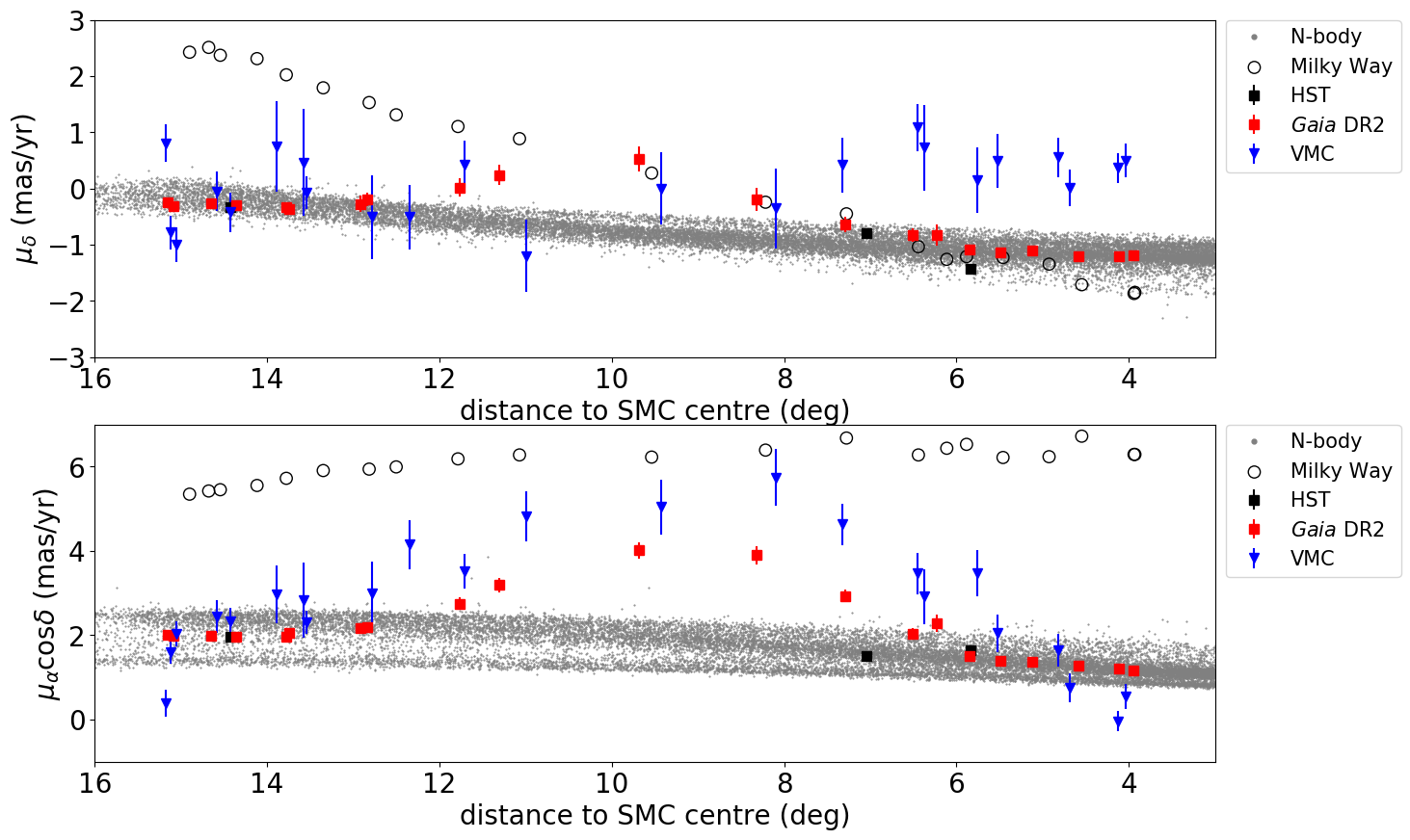}
   \caption{Proper motion in $\mu_{\delta}$ and $\mu_{\alpha}\cos{\delta}$ as a function of projected distance from the SMC centre. Grey dots in the background indicate the distribution of the simulated particles \citep{Diaz2012}. Blue triangles show the median VMC proper motion, while the red squares show the median $Gaia$~DR2 proper motion of stars across the Magellanic Bridge. Black diamonds indicate HST proper motion measurements \citep{Zivick2019}. Simulated proper motions (GalMod) of the Milky Way foreground stars are shown as black circles.
   }
   \label{PMdec}%[ht]
    \end{figure*}

\begin{figure*}
   \centering
   \includegraphics[width=\textwidth]{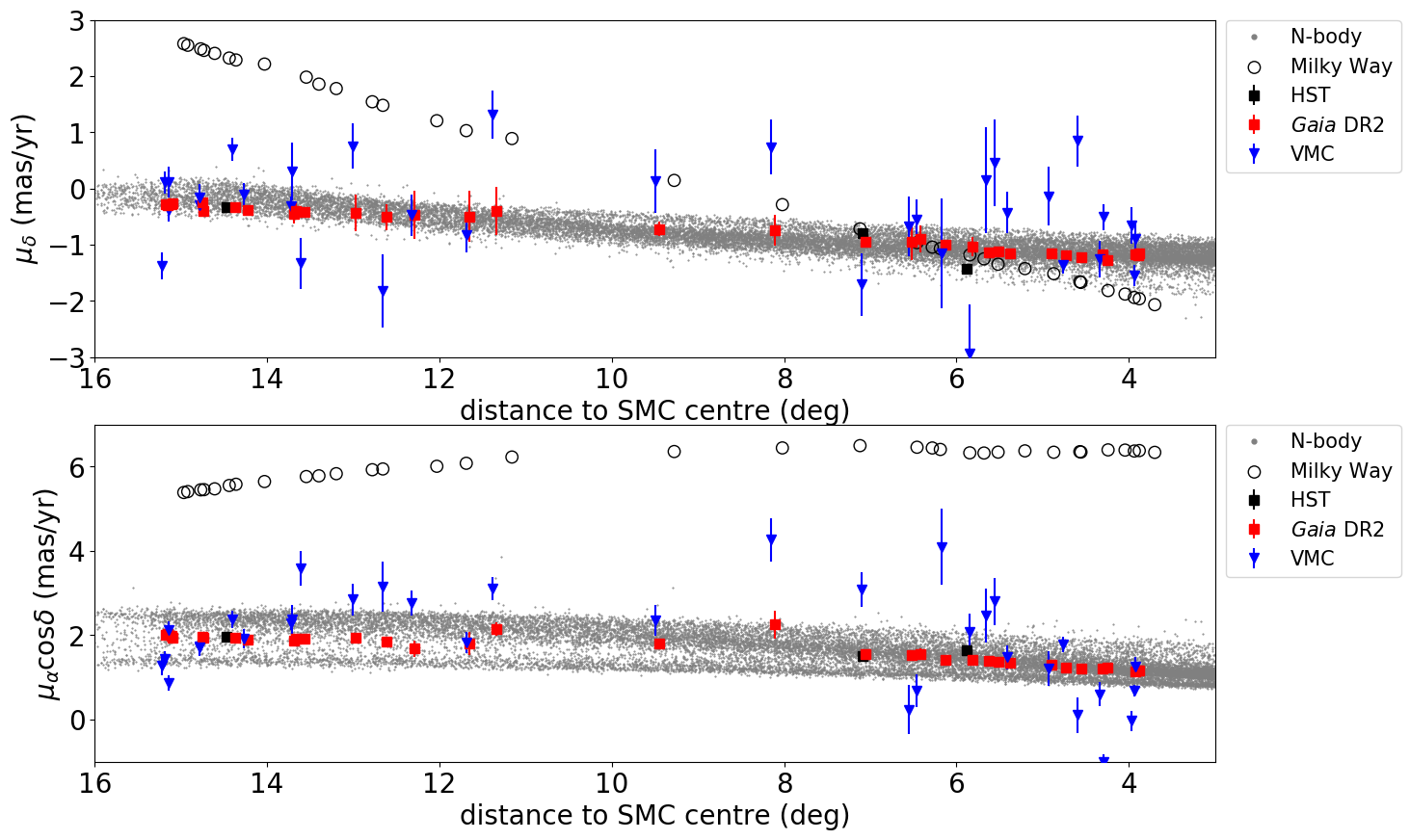}
   \caption{Proper motions $\mu_{\delta}$ and $\mu_{\alpha}\cos{\delta}$ for the sample of stars selected based on StarHorse distances (>30~kpc). Proper motions of VMC (blue triangles), $Gaia$~DR2 (red squares), $N$-body particles (grey dots), HST measurements (black squares) and the GalMod Milky Way foreground (empty circles) are shown as a function of projected distance from the SMC centre.}
              \label{PMSH}%
   \end{figure*}

%13.1875, -72.82861111
To study the proper motion of the Magellanic Bridge and to compare it with previous determinations we calculated the angular distance of each source to the centre of the SMC~(13.19$^\circ$~in~right ascension,~$-$72.83$^\circ$~in~declination; \citealt{Crowl2001}). This was done for the sources within each sample and the particles of the $N$-body simulation (see Sect.~\ref{Analysis}). Fig.~\ref{PMdec} shows proper motions in $\mu_{\delta}$ (top) and $\mu_{\alpha}\cos{\delta}$ (bottom) as a function of projected distance from the SMC centre for the Magellanic Bridge stars selected by method~1 (Sect.~\ref{method1}). It compares the median VMC proper motions (blue), the median $Gaia$~DR2 proper motions (blue) and the $N$-body simulation (grey). Structures within the model are likely caused by stripping material with different angular momenta because the simulation is collisionless. Such features are expected to be less visible in the observations, but they are shown here to indicate a range of possible values. Recent HST proper motion measurements \citep{Zivick2019} shown in black fall well within the predicted spread of the model. Both VMC and $Gaia$~DR2 proper motions are similar to those predicted by the model in denser regions on the LMC side of the Magellanic Bridge (left), but show a discrepancy towards larger proper motions in the central regions. An increasing ratio of MW foreground stars could cause this discrepancy since VMC and $Gaia$~DR2 measurements are more aligned towards the median proper motion of the MW foreground stars. The two HST measurements closer to the SMC seem also to align better with the median MW foreground proper motions, but only in the $\mu_{\delta}$ (top) measurement, where a separation between Magellanic Bridge and MW stars based on proper motions is less clear compared to the proper motion in $\mu_{\alpha}\cos{\delta}$. The VMC proper motions also show a discrepancy with respect to the model on the SMC side of the Magellanic Bridge despite a high source density. The VMC proper motions close to the SMC align neither with the $Gaia$~DR2 nor with the median MW foreground proper motions. However they show similar trends as in the proper motion maps presented by \citet{Murray2019}, especially for the proper motions towards the north (also visible in the top panel of Fig.~\ref{PM_map_method1}). The differences could be due to stellar populations behind the Magellanic Bridge, since the SMC is also known to have a significant depth (i.e. $\sim$14~kpc, \citealt{Subramanian2012}; 10$-$23~kpc, \citealt{Nidever2013}), but this depends on the stellar tracers and for some tracers there is hardly any depth \citep{deGrijs2015}. A similar proper motion trend was found in the SMC centre by \citet{Niederhofer2018b}, where the proper motions of stars in the regions of the highest stellar densities did not agree with those of the nearby regions. The discrepancy on the SMC side of the Magellanic Bridge still remains when using the much cleaner StarHorse sample (Fig.~\ref{PMSH} top).
The StarHorse sample shown in Fig.~\ref{PMSH} represents a very clean sample of Magellanic Bridge stars. Both the median VMC (blue) and $Gaia$~DR2 (red) proper motions seem to be less contaminated by the MW foreground stars, especially in $\mu_{\alpha}\cos{\delta}$ (bottom). Both VMC and $Gaia$~DR2 proper motions show a flow of stars from the SMC to the LMC, which supports simulations of the stripping of the SMC resulting from its dynamical interaction with the LMC. This was also found by \cite{Zivick2019}, using HST and $Gaia$~DR2 data and their results support the $N$-body simulation by \citet{Diaz2012}. Our sample, based on StarHorse distances and $Gaia$~DR2 proper motions, shows the best agreement yet between measurement and simulation. Minor discrepancies occur mainly in regions of low stellar density, those affected by MW foreground stars, and this effect is larger in $\mu_{\alpha}\cos{\delta}$ compared with $\mu_{\delta}$. The VMC sample is the most influenced by this effect due to the fact that $Gaia$~DR2 parallaxes for sources with $G>18$~mag are less reliable and fewer sources are available with parallax measurements. This suggests that a large number of MW foreground stars are still contaminating the VMC sample. Excluding stars without $Gaia$~DR2 parallax measurements removed to many stars from the VMC sample such that the required numbers to calculate reliable proper motion medians were not met.

An additional epoch extending the VMC time baseline will improve on this issue in the future as explored by \citet{Niederhofer2018b}. Some of the inconsistency between simulated and observed proper motions could also improve with newer models. Further improvements will also be achieved by changing the input to the astrometric solution from the 2MASS to $Gaia$~DR2, this would reduce systematic uncertainties related to 2MASS. Improving the VMC proper motion measurements is desirable to have measurements independent of $Gaia$~DR2 proper motions. At present, $Gaia$~DR2 is strongly limited by crowding in the central regions of the MCs and the VMC survey reaches in general fainter sources. $Gaia$~DR2 and VMC proper motions agree within the uncertainties, but discrepancies are visible in specific regions. VMC proper motions show the most significant discrepancy with respect to the model and $Gaia$~DR2 closer to the SMC, while they align well on the LMC side of the Magellanic Bridge. Both methods also enable us to separate individual populations (e.g. main sequence and RGB stars) since they both significantly increase the sample size (see Appendix A). The simulation fits the RGB star population (boxes C1 and C2) better than the young main sequence in dense regions, mainly due to a smaller number of main sequence stars in the sample especially on the LMC side. However the main sequence stars selected based on method~1 show a very clean sample, visible in their median $Gaia$~DR2 proper motions. There are not enough main sequence stars for reliable VMC proper motions, but they show similar trends as the corresponding RGB stars (Fig.~A.1).

\section{Summary and conclusion} \label{Conclusion}
We have analysed near--infrared data from the VMC survey of a large area of the Magellanic Bridge (23.01~deg$^{2}$) to gain a better understanding of its formation and of the most recent interaction between the LMC and SMC. We derived stellar proper motions within 13 tiles using multi-epoch K$_{s}$-band observations across a time baseline of 635-1280 days, depending on the tile. We tested two methods of reducing the influence of MW foreground stars on our proper motion measurements. 
The first method, a combination of $Gaia$~DR2 parallaxes and VMC-CMD selection criteria, proved to be very efficient in removing MW foreground stars. When comparing our results with an $N$-body simulation \citep{Diaz2012} and recent HST measurements \citep{Zivick2019} we confirmed a bulk motion of stars from the SMC towards the LMC, which was first shown by \citet{Schmidt2018}. A significant discrepancy of the VMC proper motions with respect to the model and $Gaia$~DR2 values close to the SMC suggests that further studies are needed to fully understand the complexity of the SMC kinematics, while the outer LMC regions of the Magellanic Bridge seem to be more regular. We found a discrepancy between model and measurements (both $Gaia$ and VMC) in the central region of the Magellanic Bridge, where the stellar density decreases, which is probably due to the influence of MW foreground stars. This shows that this first method to remove MW foreground stars is mainly applicable to dense stellar regions, where the significant increase of sources leads to a high spatial resolution. The second method, using StarHorse distances, allowed us to obtain the cleanest sample of Magellanic Bridge stars and to derive the first reliable stellar proper motion measurement of the central region of the Magellanic Bridge. We obtained median proper motions of 1.80$\pm$0.25~mas~yr$^{-1}$ in right ascension and $-$0.72$\pm$0.13~mas~yr$^{-1}$ in declination.
The current accuracy of the measurements is limited by the challenge of isolating stars associated with the Magellanic Bridge from those of the MW foreground, this is mainly due to the limitations of StarHorse. An additional way to remove MW foreground stars would be to use radial velocities. However, there are too few measurements of radial velocities of stars in the Magellanic Bridge. $Gaia$~DR2 radial velocities are currently limited to a small number of bright stars (G$<$13~mag), but in the future this will be rectified with observations with the 4-metre Multi-Object Spectrograph-Telescope (4MOST), see \citet{4mostCioni2019} for details. We also found that the Magellanic Bridge is stretching because the residual motions of the two opposite sides of the Magellanic Bridge are clearly moving apart relative to the proper motion of the central region. This is consistent with the model and shows that the Bridge stars on the LMC side merge into the LMC disc, while stars on the SMC side hint at a motion along with the SMC. Better proper motions and additional VMC Bridge tiles are needed to explain these complex kinematics. 
The approaches to remove MW foreground stars presented in this study are promising. There is also the opportunity to significantly increase the sample size of reliable Magellanic Bridge stars by applying StarHorse to VMC sources with $G>18$~mag, since the VMC survey detects in general fainter sources than $Gaia$ and therefore it contains more sources in total. We plan to improve the spatial resolution of our proper motion maps by developing further the methods presented in this study, including also an additional VMC epoch for all of the Bridge tiles.

%and calculating within a 2D-Voronoi-Binning median proper motions throughout the Magellanic Bridge. 

\begin{acknowledgements} We are grateful to the anonymous referee for the useful
comments and suggestions. This project has received funding from the European Research Council (ERC) under the European Union’s Horizon 2020 research and innovation programme (grant agreement no. 682115). We thank the Cambridge Astronomy Survey Unit (CASU) and the Wide Field Astronomy Unit (WFAU) in Edinburgh for providing the necessary data products under the support of the Science and Technology Facility Council (STFC) in the UK. This project has made extensive use of the Tool for OPerations on Catalogues And Tables (TOPCAT) software package (Taylor 2005) as well as the following open-source Python packages: Astropy (The Astropy Collaboration et al.2018), matplotlib (Hunter 2007), NumPy (Oliphant 2015), Numba, and Pyraf. \end{acknowledgements}

\bibliographystyle{aa} % style aa.bst
\bibliography{bibtex.bib}
\newpage
\begin{appendix}
\section{Separating RGB and main sequence stars}\label{apendix}
 Fig.~\ref{apendix} shows the proper motion measurements after removing MW foreground stars using method~1 (see Sect.~\ref{method1}) divided into two populations based on the CMD selection shown in Fig.~\ref{BRI_full_cmd}. The top panels show RGB stars (boxes C1 and C2) and the bottom panels show main sequence stars (box A). The RGB stars show the same trends as in Fig.~\ref{PMdec}, since they represent the majority of stars. The main sequence stars, although being a much smaller sample, appear as a clean sample similar to the one obtained removing MW foreground stars using method~2 (see Sect.~\ref{method2} and Fig.~\ref{PMSH}). The discrepancies between the $Gaia$~DR2 and VMC proper motions close to the SMC are only present with the RGB stars. 

\begin{figure*}
   \centering
   \includegraphics[width=0.9\textwidth]{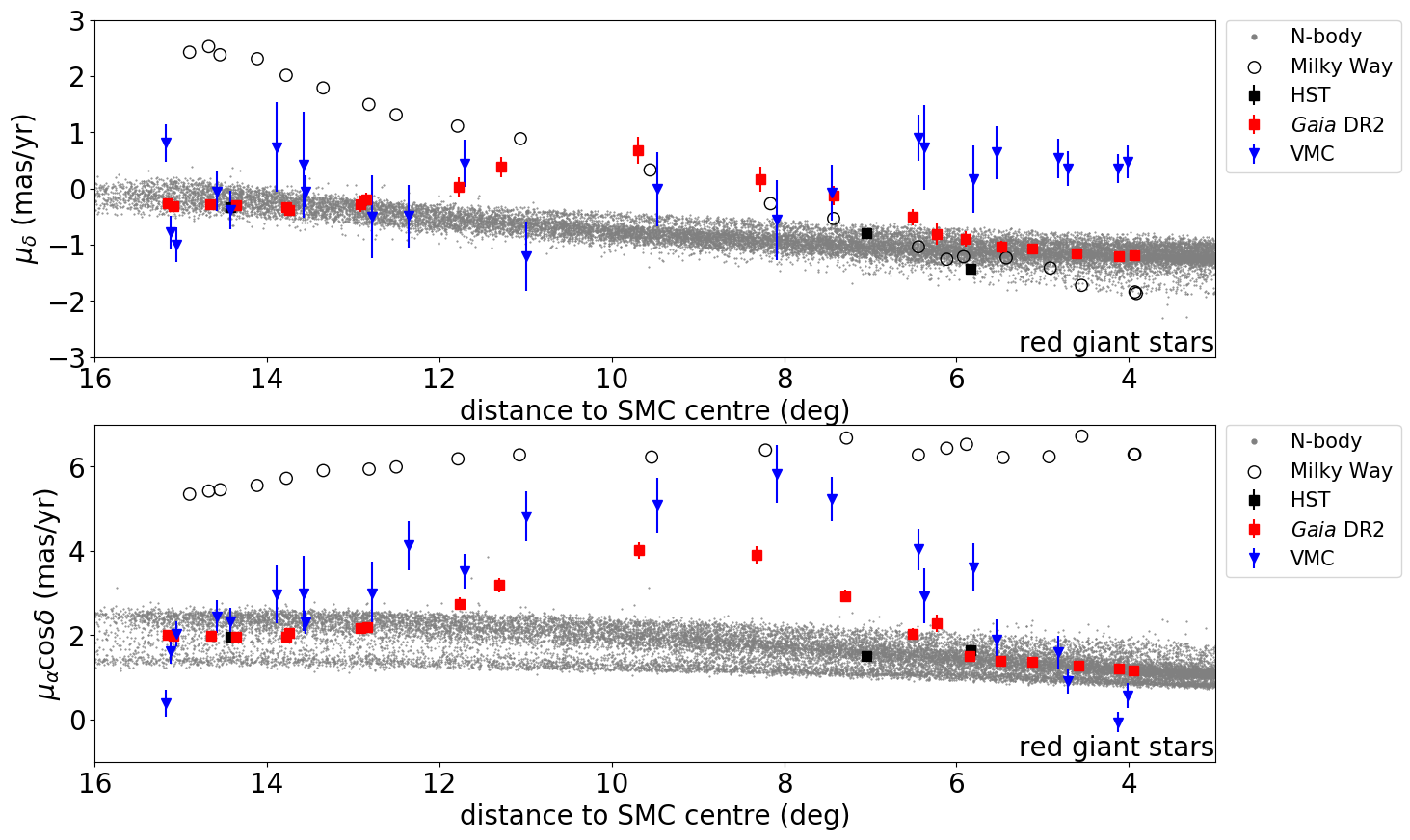}
   \includegraphics[width=0.9\textwidth]{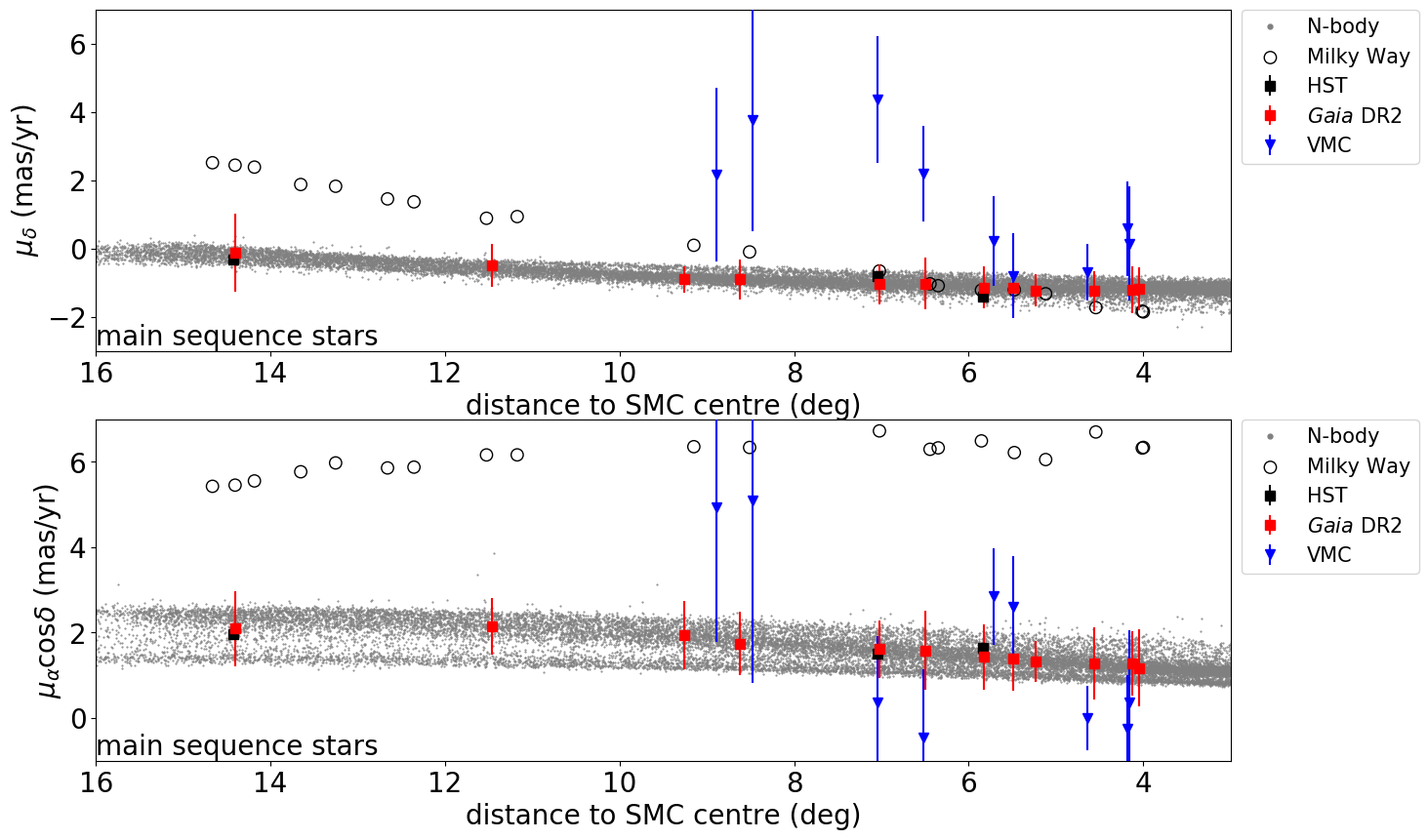}
   \caption{Same as Fig.~\ref{PMdec}, but separated in RGB stars (boxes C1 and C2) and main sequence stars (box A).}
              \label{apendix}%
   \end{figure*}

\end{appendix}

%\begin{appendix} %First online appendix
%\section{Control plots}
%Extra plots not added jet
%\end{appendix}
%-----------------------------------------------------------------------
\end{document}